 \documentclass{ajour}
\usepackage{epsfig}
\begin{document}

%







\newfont{\fraktur}{eufm10 at 11pt}
\newfont{\fields}{msbm10 at 10pt}
\newfont{\fieldabstract}{msbm10 at 8pt}
\newfont{\bigscript}{eusm10 at 10pt}
\newfont{\scriptcaption}{eusm10 at 8pt}

\def\hf{\textstyle{1\over2}}
\def\half{{1\over2}}

\def\beq{\begin{equation}}
\def\eeq{\end{equation}}
\def\beqa{\begin{eqnarray}}
\def\eeqa{\end{eqnarray}}

\title{Gauge Theory of Riemann Ellipsoids}


\author{G. Rosensteel and J. Troupe}
\affil{Department of Physics, Tulane University,
New Orleans, LA 70118, and \\
Institute for Nuclear Theory, University of Washington,
Seattle, WA 98195}

\email{grosens@mailhost.tcs.tulane.edu; jtroupe@mailhost.tcs.tulane.edu}

\authorrunninghead{G. Rosensteel and J. Troupe}
\titlerunninghead{Gauge Theory of Riemann Ellipsoids}

\abstract{The classical theory of Riemann ellipsoids is formulated naturally as a gauge theory 
based on a principal $G$-bundle ${\cal P}$. The structure
group $G=SO(3)$ is the vorticity group, and the bundle ${\cal P}=GL_+(3,\mbox{\fieldabstract R})$ is 
the connected component of the general linear group. The base
manifold is the space of positive-definite real $3\times 3$ symmetric
matrices, identified geometrically with the space of inertia ellipsoids.
The bundle ${\cal P}$ is also a Riemannian manifold whose metric is determined by the
kinetic energy. Nonholonomic constraints determine connections on the bundle.
In particular, the trivial connection corresponds to rigid body motion, the natural Riemannian connection - to irrotational flow, and the invariant connection - to the falling cat. The curvature form determines the fluid's field tensor which is an analogue of the familiar Faraday tensor. 
}


\begin{article}








\section{INTRODUCTION}
A Riemann ellipsoid is a constant density fluid bounded by an ellipsoidal
surface and constrained to a linear velocity field.  In addition to rotational
and vibrational motion, Riemann ellipsoid dynamics admits vorticity degrees
of freedom. The vorticity allows the moment of
inertia to be varied continuously between its rigid body and its irrotational
fluid limits. It is this flexibility that 
enables the theory to model diverse rotating systems \cite{RIE60, LEB, Chandrasekhar69, DYS68, Rose88}.  

An ellipsoid is defined by a positive-definite real symmetric $3\times 3$ 
matrix $q$; the points $X=(X_1,X_2,X_3)$ of its surface in Euclidean space
are solutions to the quadratic equation
\beq
\sum_{ij} q_{ij}^{-1} X_i X_j = 1 . 
\eeq
In particular, when $q = \mbox{diag}(a_1^2, a_2^2, a_3^2)$, the surface
is an ellipsoid with major axis half-lengths equal to $a_1, a_2, a_3$ and the
ellipsoid's principal axes are aligned with the Cartesian axes. The space of all ellipsoidal surfaces is identified with the six dimensional manifold of such
matrices $q$:
\beq
Q = \{ q\in M_3(\mbox{\fields R}) \ | \ q^t = q, q>0 \} .
\eeq

The configuration space for Riemann ellipsoids is the general linear group
\beq
{\cal P} = \{ \xi \in M_3(\mbox{\fields R}) \ |\ \det \xi > 0 \} = GL_+(3,\mbox{\fields R}) ,
\eeq
because only linear motions are allowed. The elegance and beauty of the theory
of Riemann ellipsoids are derived ultimately from the group structure on the
configuration space ${\cal P}$. 

The general linear group transforms an ellipsoid
into another ellipsoid that is deformed and rotated. If $q\in Q$ and $\xi \in {\cal P}$, then this spatial transformation corresponds to $q\longmapsto \xi q \xi^t$. In particular, the unit sphere with $q=I$ is changed into the ellipsoid with $q=\xi \xi^t$. 
Thus, the natural mapping $\pi$ from the group ${\cal P}$ onto the ellipsoidal space $Q$
\beqa
\pi : {\cal P} & \longrightarrow & Q \nonumber \\
         \xi & \longmapsto & q = \xi \xi^t 
\eeqa
is surjective (onto).

The group ${\cal P}$ also acts on itself. The group elements can be multiplied in two ways, on the left or on the right. The left and right actions are distinct when the group is not Abelian. They  define two different transformations of the group on itself. Left multiplication of elements $\xi\in {\cal P}$ by $r\in SO(3)$
\beq
\xi \longmapsto L_{\! r} \xi = r \xi 
\eeq
corresponds to physical rotations $r$ in three-dimensional Euclidean space. In contrast, right multiplication of $\xi\in {\cal P}$ by $g\in G = SO(3)$,
\beq
\xi \longmapsto R_g \xi = \xi g^{-1},
\eeq
corresponds to vortex motion. Although the groups $G$ and $SO(3)$ are mathematically isomorphic, their distinct actions on ${\cal P}$ imply different physical interpretations.

With respect to left multiplication by the rotation group element $r\in SO(3)$, the ellipsoid with inertia tensor $q = \pi(\xi) = \xi \xi^t$ is transformed into the rotated ellipsoid with inertia tensor $\pi(L_{\! r} \xi) = r \xi \xi^t r^t = r q r^t$. But right multiplication by the vorticity group element $g \in G$ leaves the inertia ellipsoid invariant, $\pi(R_g \xi) = \xi g^{-1} g \xi^t  = q$, since $g^{-1} = g^t$. This invariance is expressed more elegantly by the composition of mappings
\beq 
\pi \circ R_g = \pi, \mbox{\ for all\ } g\in G.
\eeq  
Hence $G$ is a group of motions internal to the ellipsoidal
surface.

Because of the right invariance of the surjective mapping $\pi$ with respect to the group $G$, the configuration space ${\cal P}$ is a principal fiber bundle over the base manifold $Q$ with structure group $G$ \cite{Steenrod, KN}. It is the purpose of this article to show that the classical theory of
Riemann ellipsoids is expressed naturally in the mathematical framework of the
bundle ${\cal P}$. The bundle ${\cal P}$ has two additional structures that are important physically. ${\cal P}$ is a Riemannian manifold whose metric is determined by the kinetic energy in three-dimensional Euclidean space. ${\cal P}$ may also be equipped with a differential structure, or connection. The nonholonomic constraints to irrotational flow and the falling cat problem correspond to the Riemannian connection and the invariant connection, respectively.

Recently Littlejohn and Reinsch \cite{LJ} and earlier Shapere and Wilczek \cite{SW} presented a gauge theory of rotating systems in which the gauge group is the rotation group, and the gauge conserved quantity is the angular momentum. This article describes a fundamentally different model in which the gauge group is the vorticity group and the gauge conserved quantity is the Kelvin circulation (the angular momentum is also conserved).

The mathematical theory of bundles is the setting for the gauge theories of
fundamental forces \cite{ORAIF, GOCK}. For the Riemann ellipsoid model, the gauge group is the vorticity group $G = SO(3)$, and, therefore, it is a non-Abelian gauge theory.
The situation is reminiscent of the Weinberg-Salam theory of the electroweak interaction \cite{Weinberg} for which the structure group is $SU(2)\times U(1)$. An important mathematical difference is that the base manifold for the electroweak problem is Minkowski space-time, while the Riemann ellipsoid base manifold is the space of ellipsoidal surfaces. Although Cartesian coordinates $(t,x,y,z)$ are convenient for space-time, a coordinate system for $Q$ requires Euler angles to describe the ellipsoid's orientation. This technical complexity can be overcome by using Cartan's method of moving frames \cite{KN}. For the rotation group $SO(3)$, the angular momenta serve as the moving frame, and, for the gauge group $G$, the Kelvin circulation is used.
In this sense, the differential
geometry of the Riemann ellipsoid theory is more complicated mathematically than the
space-time based theories of fundamental physics. 

In $\S 2$ the theory of rotating triaxial rigid bodies is presented using group theoretic and geometrical methods that generalize to the Riemann ellipsoid problem. The rigid body configuration space is $SO(3)$ that is both a Lie group and a Riemannian manifold. Applying the Cartan method of moving frames, the velocity of a rotating body is expressed in terms of the angular momentum which is an invariant vector field. The kinetic energy determines a metric on the rotation group. The Euler rigid body equations are obtained from Lagrange's equations.

The Riemann ellipsoid theory is presented in $\S 3$. The interrelated Lie group, bundle,  and Riemannian structures of the configuration space ${\cal P}$ are reviewed and the equations of motion are derived in the Lagrange formalism. $\S 4$ is the heart of the paper. Here the differential geometry of bundle connections is used to show that the nonholonomic constraints to irrotational flow and the falling cat are realized by mathematically natural connections.

\section{RIGID BODY ROTATION}
Although the rigid body problem is solved in basic mechanics textbooks, the method of solution does not generalize easily to higher dimensional non-Euclidean configuration manifolds, e.g., the Riemann ellipsoid space ${\cal P}$. The source of the limitation in the conventional technique is the use of Euler angle coordinates for the configuration manifold $SO(3)$. In this section the rigid body problem is solved using the method of moving frames. Cartan's more powerful technique enables the solution of mechanics problems in higher dimensional non-Euclidean spaces.

In the conventional solution of the rigid body problem, Euler angles provide coordinates for the configuration space {\em and} a basis for each tangent space. For the rigid body problem, the natural physical and geometrical choice for a tangent space basis consists of the angular momentum vectors. No coordinate system can define directly the angular momenta because their Lie brackets do not vanish. Cartan's moving frames provide each tangent space with bases that vary from point to point and cannot be expressed consistently in a fixed coordinate chart.

\subsection{Tangent Vectors} 
A vector $V_{\! R}$ tangent to the group manifold $SO(3)$ at the point $R$ determines
a unique directional derivative as follows: if   
$f$ is a real-valued differentiable function defined 
on an open neighborhood of the point $R$ 
in the rotation
group, then the real number
$(V_{\! R} f)(R)$ is defined as the derivative of $f$ in the direction
$V_{\! R}$ evaluated at 
$R\in SO(3)$. A vector field $V$ on $SO(3)$ is defined by a smooth assignment of a
directional derivative $V_{\! R}$ to every point $R\in SO(3)$. Hence a vector
field $V$ may be regarded as a first order differential operator on $SO(3)$:
if $f$ is a smooth function on $SO(3)$, then $Vf$ is also
a smooth function defined by $(Vf)(R) = (V_{\! R} f)(R)$, for all $R\in SO(3)$.
The space of all vector fields on $SO(3)$ is identified with the linear derivations, i.e., the space of all mappings $V$ from the space of smooth functions on $SO(3)$ into itself that satisfy the linearity and Leibniz conditions,
\beqa
V (a f + b g) & = & a (Vf) + b (Vg)  \nonumber \\
V (f g)  & = & f (Vg) + (Vf) g ,
\eeqa
where $f$ and $g$ are smooth functions on $SO(3)$ and $a,b$ are real numbers. 

The Lie bracket of two vector fields is the commutator of their corresponding derivations. Since it is a first 
order differential operator, the bracket of two vector fields is a vector field.
In this way, the space of all vector fields is an infinite-dimensional Lie algebra. 
The components of a vector field depend on the choice of a coordinate system for
$SO(3)$. If Euler angles $(\theta^1, \theta^2, \theta^3)$ are taken, then the
components of a vector field $V$ are given by applying $V$ to the coordinate
functions, $V(\theta^i) = V^i$. Expanded into components, the vector field is
\beq
V = \sum_i V^i \frac{\partial}{\partial \theta^i} , 
\eeq
because $\partial \theta^j /\partial \theta^i = \delta^j_i$. 

There are two 
principal advantages to regarding vector fields as first order differential 
operators. The first is that the Lie bracket may be defined easily. The
second is that the differential operator $V$ is independent of the 
coordinate system, although its components are coordinate dependent. If a second coordinate system is chosen for $SO(3)$, say
$(\psi^1,\psi^2,\psi^3)$, then the vector field components change to $V(\psi^i) = v^{i}$.
But the differential operator, expressed in the new coordinate chart, stays
the same,
\beq
V = \sum_i v^i \frac{\partial}{\partial \psi^i},  
\eeq  
since the components of a vector transform as the Jacobian, $v^i ( \partial \theta^j 
/ \partial \psi^i ) = V^j$. 

It is a tactical error to work directly with Euler angles or any other non-Euclidean coordinate system. Even if results
are obtained successfully, the derivations are unilluminating, and can make
trivial results seem very complicated, c.f., Goldstein's derivation of the
Euler rigid body equations \cite{Goldstein}. Instead it is best to use consistently a Euclidean 
coordinate system. Consider the nine-dimensional space $M_3(\mbox{\fields R})$ of all
$3\times 3$ real matrices. The rotation group $SO(3)$ is viewed as a closed,
bounded three-dimensional surface in $M_3(\mbox{\fields R})$. The nine entries of each matrix provide a convenient global coordinate system for $M_3(\mbox{\fields R})$. In terms of this coordinate system, the points of the $SO(3)$ surface satisfy two polynomial conditions that
are naturally expressed in matrix notation
\beq
SO(3) = \{ R\in M_3(\mbox{\fields R}) \ | \ R^t R = I,\, \det R=1 \} \, .
\eeq 
There are six independent quadratic polynomial conditions on the $9$ matrix
coordinates $R_{ij}$, viz., $\sum_k R_{ki} R_{kj} = \delta_{ij}$, for $i\leq j$.
Hence the surface $SO(3)$ is $9-6=3$ dimensional.  
When a rigid body is rotating, its motion defines a smooth curve
$t\longmapsto R(t)$ in the surface of $SO(3)$. The tangent or velocity vector to this curve in the surface $SO(3)$ is given in the matrix coordinate chart of $M_3(\mbox{\fields R})$ by
\beq
V(t) = \sum_{ij} \dot{R}_{ij}(t) \frac{\partial }{\partial R_{ij}} \, . 
\eeq

\subsection{Left and Right Invariant Vector Fields}
Among all possible vector fields, the invariant ones are singled out for
their mathematical simplicity and physical relevance. In the following discussion, the left invariant vector field at a point $R(t)$ is shown to correspond to the angular velocity in the laboratory (inertial) frame, and the right invariant vector field at a point $R(t)$ is shown to correspond to the angular velocity in the body-fixed (intrinsic) frame.

For each element of 
the Lie algebra of the rotation group, a left invariant vector field and 
a right invariant vector field can be defined. Let $\mbox{\fraktur so}(3)$ denote
the Lie algebra of the rotation group $SO(3)$,
\beq
\mbox{\fraktur so}(3) = \{ \Omega \in M_3(\mbox{\fields R}) \ | \ \Omega^t = -\Omega \} .
\eeq
As it is customary, the upper case $SO(3)$ denotes the group while the lower
case $\mbox{\fraktur so}(3)$ denotes the algebra.
The rotation of a rigid body $t\longmapsto R(t)$ naturally defines a Lie algebra
element at each point. Since $R^t R=I$ at each time t, its vanishing
derivative with respect to $t$ implies that the matrix
\beq
\Omega^L (t) = R^t \dot{R}
\eeq 
is antisymmetric. $\Omega^L$ has the physical interpretation of the angular
velocity in the laboratory (inertial) frame. In this article superscript $L$ will indicate a quantity in the laboratory (inertial) frame.
 
A basis for the three dimensional vector space $\mbox{\fraktur so}(3)$ is the set of
antisymmetric matrices $\mbox{\fraktur e}_i$ for $i=1,2,3$,
with matrix elements given by $(\mbox{\fraktur e}_i)_{jk} \equiv \varepsilon_{ijk}$.
If $\Omega^L\in \mbox{\fraktur so}(3)$ is expanded in this basis, $\Omega^L = \sum \omega_i^L
\mbox{\fraktur e}_i$, $\mathbf{\bbf{\omega}}^L$ is interpreted as the laboratory
frame angular velocity vector.

When $\Omega \in \mbox{\fraktur so}(3)$ is independent of time, define $(\mbox{\bigscript L}_\Omega)_e$ to be the tangent to the curve
$\theta \longmapsto \exp(\theta \Omega)$ in the group $SO(3)$ that passes through
the group identity $e$ when $\theta=0$. For example a rotation about the 3-axis is
given by
\beq
\exp(\theta \mbox{\fraktur e}_3) = \left( \begin{array}{ccc} 
\cos \theta & \sin \theta & 0 \\
-\sin \theta & \cos \theta & 0 \\
0 & 0 & 1 \end{array}
\right) . 
\eeq 
A left invariant vector field is defined by spreading this tangent vector
around the group manifold. At the point $R\in SO(3)$, define $(\mbox{\bigscript L}_\Omega)_{\! R}$
 to be the tangent to the curve $\theta \longmapsto R \cdot \exp(\theta\Omega)$ that passes through $R\in SO(3)$. 
Similarly a right invariant vector field $(\mbox{\bigscript R}_\Omega)_{\! R}$ is the tangent to the curve 
$\theta\longmapsto \exp(-\theta\Omega) \cdot R$, see Figure 1.

As first order differential operators on $SO(3)$, the
invariant vector fields are given by
\beqa
(\mbox{\bigscript L}_\Omega)_{\! R} & = & (R \Omega)_{ij}\frac{\partial}{\partial R_{ij}} \mathletter{a} \\
(\mbox{\bigscript R}_\Omega)_{\! R} & = & -(\Omega R)_{ij}\frac{\partial}{\partial R_{ij}} , \mathletter{b} \label{diffop}
\eeqa
for each $\Omega\in \mbox{\fraktur so}(3)$.  These derivations are called Lie derivatives.

The two vector fields are tangent to the three dimensional surface of
orthogonal unit determinant matrices contained within the nine dimensional
manifold of all $3\times 3$ real matrices. The velocity 
vector of a curve $t\longmapsto R(t)$ may be expressed at each instant in time
as the value of left invariant vector field at the point $R(t)$
\beq
V(t) =  \sum_{ij} \dot{R}_{ij}(t) \frac{\partial }{\partial R_{ij}} = \sum_{ij} (R \Omega^L(t))_{ij} \frac{\partial }{\partial R_{ij}} = ( \mbox{\bigscript L}_{\Omega^L(t)})_{R(t)} \, .
\eeq

As $\Omega$ ranges over the basis $\mbox{\fraktur e}_i$ of $\mbox{\fraktur so}(3)$, the left and right 
invariant vector fields provide separate bases for each tangent space
to the group manifold $SO(3)$. To see the relation between the two invariant bases,
note that the curve whose tangent at $R$ is a left invariant vector field
may be expressed alternatively as
\beqa
\theta & \longmapsto & R\cdot \exp(\theta \Omega) \nonumber \\
& = & R \exp(\theta\Omega) R^{-1} R  \nonumber \\
& = &  \exp(\theta Ad_{\! R} \Omega) R \, ,
\eeqa
which is a curve tangent to a right invariant vector field: 
\beq 
(\mbox{\bigscript L}_\Omega)_{\! R} = - (\mbox{\bigscript R}_{Ad_{\! R}\Omega})_{\! R} \, . \label{lefttoright} 
\eeq 
Here the adjoint transformation $Ad$ is introduced: For each $R\in SO(3)$ and
$\Omega\in \mbox{\fraktur so}(3)$, define the Lie algebra element Ad$_{R}\Omega = 
R\Omega R^{-1}$ in $\mbox{\fraktur so}(3)$. The adjoint mapping $\Omega \longmapsto Ad_{\! R} \Omega$
is equivalent to ordinary rotation of the corresponding velocity vector $\mathbf{\bbf{\omega}}
\longmapsto R \mathbf{\bbf{\omega}}$. The equivalence depends on the determinant
of a special orthogonal matrix being equal to $+1$, and, hence, $\mathbf{\bbf{\omega}}$
is an axial vector.

If $\Omega^L(t)$ is the laboratory frame angular velocity, $\Omega(t) = 
\mbox{Ad}_{R(t)}\Omega^L(t)= \dot{R} R^t$ is the angular velocity in the rotating
frame. Hence, the velocity in the laboratory frame is expressed in
the rotating frame by
\beq
V(t) = (\mbox{\bigscript L}_{\Omega^L(t)})_{R(t)} = -(\mbox{\bigscript R}_{\Omega(t)})_{R(t)} .
\eeq 
When $\Omega = \sum_i \omega_i \mbox{\fraktur e}_i$ is expanded in the usual basis for $\mbox{\fraktur so}(3)$, the vector $\mathbf{\bbf{\omega}} = R \mathbf{\bbf{\omega}}^L$ is called the angular velocity vector 
in the rotating frame.

The left and right invariant vector fields are special for various reasons:

\noindent{\bf 1.} Each is a representation of the Lie algebra, i.e., 
\beqa
\mbox{\bigscript L}_{[\Omega_1,\Omega_2]} & = & [\mbox{\bigscript L}_{\Omega_1}, \mbox{\bigscript L}_{\Omega_2}] \mathletter{a} \\
\mbox{\bigscript R}_{[\Omega_1,\Omega_2]} & = & [\mbox{\bigscript R}_{\Omega_1}, \mbox{\bigscript R}_{\Omega_2}] .  \mathletter{b} 
\eeqa
The invariant vector fields are finite-dimensional subalgebras of the
algebra of all vector fields. Both the left and the right invariant vector fields
are faithful, but reducible, representation of the Lie algebra $\mbox{\fraktur so}(3)$. With
respect to the Haar measure, each $\mbox{\bigscript L}_\Omega$ and $\mbox{\bigscript R}_\Omega$ is a skew-adjoint operator on
the space of square-integrable functions on $SO(3)$. In particular, the 
Hermitian operators
\beqa
\hat{L}_j & \equiv & -i\, \mbox{\bigscript L}_{\mbox{\fraktur e}_j} \nonumber \\
\hat{I}_j & \equiv & i\, \mbox{\bigscript R}_{\mbox{\fraktur e}_j}
\eeqa
are the quantum mechanical angular momenta in the laboratory (inertial)
frame and the body-fixed (intrinsic) frame, respectively.

\noindent{\bf 2.} As Lie derivatives, they are invariant with respect to the group 
multiplication. For each $g$ in the $SO(3)$, let $L_g$ and $R_g$ 
denote left and right multiplication,
\beqa
L_g (R) & = & g R  \mathletter{a}  \\
R_g (R) & = & R g^{-1} ,  \mathletter{b} 
\eeqa
for all $R$ in $SO(3)$.
Both multiplications are group homomorphisms, e.g., $L_g L_h = L_{gh}$ for all
$g, h$ in the group. The group homomorphism $L_g$ maps a neighborhood of $R\in
SO(3)$ to a neighborhood of $g R \in SO(3)$. The curve 
$\theta \longmapsto R \exp(\theta \Omega)$
through $R$ is mapped by $L_g$ to the curve $\theta \longmapsto g R \exp(\theta \Omega)$ through $g R$. Hence the left group multiplication induces a mapping, 
denoted by $(L_g)_{*}$, from
the tangent space at $R$ to the tangent space at $L_g(R) = g R$. But the
tangent vector $(L_g)_*(\mbox{\bigscript L}_\Omega)_{\! R}$ is none other that 
$(\mbox{\bigscript L}_{\Omega})_{gR}$.
In this sense the left invariant vector field is invariant with respect to the
left group multiplication. Similarly, a right invariant
vector field is invariant with respect to right group multiplication
\beqa
(L_g)_*( \mbox{\bigscript L}_\Omega)_{\! R} & = & (\mbox{\bigscript L}_\Omega)_{gR}  \mathletter{a}   \\
(R_g)_*( \mbox{\bigscript R}_\Omega)_{\! R} & = & (\mbox{\bigscript R}_\Omega)_{Rg^{-1}} \, .  \mathletter{b} 
\eeqa
Unless a rigid body is rotating with a constant angular velocity $\Omega^L$, 
the velocity $V(t)$ will not be left invariant along the trajectory 
$t \longmapsto R(t)$. Although at each instant in time the velocity vector along
a trajectory may
be expressed as the left invariant vector $\mbox{\bigscript L}_{\Omega^L(t)}$, 
it keeps changing with time. A similar remark applies to the right invariant
vector $\mbox{\bigscript R}_{\Omega(t)}$.
\newline  
\noindent{\bf 3.} What happens when $(R_g)_*$ is applied to a left invariant vector field?
In this case, the curve $\theta \longmapsto R \exp(\theta\Omega)$ through $R$ is mapped by $R_g$ to the curve
$\theta \longmapsto R \exp(\theta\Omega) g^{-1}$ through $R g^{-1}$. But
\beqa
R \exp(\theta \Omega) g^{-1} & = & R (g^{-1} g) \exp(\theta\Omega) g^{-1} \nonumber \\
& = & (R g^{-1}) \exp(\theta\, g \Omega g^{-1}) \nonumber \\
& = & (R g^{-1}) \exp(\theta\, \mbox{Ad}_{g} \Omega ) . 
\eeqa
Therefore $(R_g)_*$ maps the left invariant vector field corresponding to 
$\Omega$ at $R$ to the left invariant vector field corresponding to Ad$_g(\Omega)$ at
$Rg^{-1}$, see Figure 2. A similar argument holds for the right invariant vector fields 
acted on by $(L_g)_*$
\beqa
(R_g)_* (\mbox{\bigscript L}_\Omega)_{\! R} & = & (\mbox{\bigscript L}_{\mbox{\tiny Ad}_g \Omega})_{Rg^{-1}} \mathletter{a} \label{adjoint1} \\
(L_g)_* (\mbox{\bigscript R}_\Omega)_{\! R} & = & (\mbox{\bigscript R}_{\mbox{\tiny Ad}_g \Omega})_{gR}. \mathletter{b} \label{adjoint2}
\eeqa 
In particular, if $g=R(t)$ is the actual rotation of a rigid body, then at
each instant
\beq 
(R_{R(t)})_* V(t) = (\mbox{\bigscript L}_{\Omega(t)})_{e} . 
\eeq

\noindent{\bf 4.} The set of left invariant vector fields $\{ \mbox{\bigscript L}_{\mbox{\fraktur e}_1} , \mbox{\bigscript L}_{\mbox{\fraktur e}_2} , \mbox{\bigscript L}_{\mbox{\fraktur e}_3} \}$ provides a basis for each tangent space. It is called a ``moving frame.'' The velocity is $V(t) = \sum \omega_i^L(t) \, (\mbox{\bigscript L}_{\mbox{\fraktur e}_i})_{\! R(t)}$. Similarly, the set of right invariant vector fields $\{ \mbox{\bigscript R}_{\mbox{\fraktur e}_1} , \mbox{\bigscript R}_{\mbox{\fraktur e}_2} , \mbox{\bigscript R}_{\mbox{\fraktur e}_3} \}$ is another moving frame and $V(t) = -\sum \omega_i(t) \, (\mbox{\bigscript R}_{\mbox{\fraktur e}_i})_{\! R(t)}$.

\subsection{Riemannian Structure}
The group manifold $SO(3)$ is a Riemannian manifold, i.e., there is a positive-definite metric defined on it. The metric depends on the moments of inertia of the rigid body and is chosen so that the rigid body's kinetic energy is proportional to the squared length of the velocity. Thus the metric on $SO(3)$ is inherited from three-dimensional Euclidean space. 

Let $\mathbf{X}_{\alpha}$ denote the position vector in an inertial frame 
for particle $\alpha$ of mass $m_\alpha$ in a system of many particles ($\alpha$ denotes the particle index). 
 Let $M =\sum m_{\alpha}$ denote the total mass of the system and let $R_0$ be a convenient unit of length. 
The dimensionless quadrupole-monopole tensor $q(X)$ for a collection of
particles is defined by  
\beq
(M R_0^2 /5)\ q(X)_{ij} = \sum_{\alpha} m_\alpha X_{\alpha i} X_{\alpha j}\, ,
\eeq 
where the sum is over the particle index $\alpha$ and $X_{\alpha i}$, $i=1,2,3$,
denotes the Cartesian components of the position vector. For a continuous fluid the
sum is replaced by an integral over the mass density distribution.

For rigid rotation there is a curve $t\longmapsto R(t)$ in the rotation group for which $R(t) \mathbf{X}_\alpha (t) = \mathbf{x}_{\alpha}$ is independent of the time $t$ for all particles $\alpha$. The motion is a collective rotation because all the particles move in an orchestrated manner according to the same linear transformation $R(t)$. As a real symmetric positive-definite matrix, $q(X)$ may be diagonalized by a rotation and its eigenvalues are real positive numbers. Hence, the reference distribution $\mathbf{x}_\alpha$ may be chosen so that its quadrupole-monopole tensor $q(x) = R q(X) R^t$ is a diagonal matrix,
\beq
(M R_0^2 /5)\ q(x)_{ij} = \sum_{\alpha} m_\alpha x_{\alpha i} x_{\alpha j} = (M R_0^2 /5) a_i^2 \delta_{ij}
\eeq 
or $q(x) = A^2 = \mbox{diag}(a_1^2, a_2^2, a_3^2)$. For a uniform density ellipsoid, $a_i$ is the half-length, in units of $R_0$, of the $i^{th}$ principal axis.

The velocity vector for each particle is $\mathbf{U}_\alpha = \dot{R}^t \mathbf{x}_\alpha$. The kinetic
energy of the system is
\beqa
T & = & \sum_{\alpha}\frac{m_\alpha}{2} \mathbf{U}_\alpha \cdot
\mathbf{U}_\alpha \nonumber \\
& = & (MR_0^2/10)\ \sum_{ij} a_i^2 \dot{R}_{ij} \dot{R}_{ij} \nonumber \\
& = & - (MR_0^2/10)\ \mbox{tr} (A^2 \Omega^2) \nonumber \\
& = &  \half( I_1 \omega_1^2 + I_2 \omega_2^2 + I_3 \omega_3^2) , 
\eeqa
where the moments of inertia of a rigid ellipsoid are $I_1 = (MR^2_0 / 5)(a_2^2 +
a_3^2)$, etc. The last line is proven using the identity $(\Omega^2)_{ij} 
= \omega_i \omega_j - \delta_{ij}\mathbf{\bbf{\omega}}\cdot\mathbf{\bbf{\omega}}$. 

At each point $R\in SO(3)$, define the metric {\bf g} as the bilinear form
\beq
{\textbf g}_{\! R}((\mbox{\bigscript R}_{\Omega_1})_{\! R},(\mbox{\bigscript R}_{\Omega_2})_{\! R}) = - {\rm tr}(\Omega_1 A^2 \Omega_2)
\eeq
for each pair of tangent vectors $(\mbox{\bigscript R}_{\Omega_1})_{\! R}, (\mbox{\bigscript R}_{\Omega_2})_{\! R}$ at $R$ where $\Omega_1, \Omega_2 \in \mbox{\fraktur so}(3)$. This metric is manifestly right invariant which is the expression of its rotational invariance in the laboratory frame.
Moreover, the rigid body kinetic energy is proportional to the squared length of its velocity
\beq
T  =  (MR_0^2/10)\ {\textbf g}_{R(t)} (V(t), V(t)) .
\eeq

\subsection{Lagrange's Equations}
The dynamical equations for the rigid body may be derived from its Lagrangian.
The Lagrangian is just the kinetic energy and the action is its integration 
along paths between two fixed points in $SO(3)$. Rigid body motion corresponds
to the path that minimizes the action with respect to all such paths. If
Euler angles are introduced, the calculation of the minimization equations
is tedious. It is better to
enlarge the space from $SO(3)$ to the 12-dimensional Euclidean configuration 
space 
whose coordinates are the nine matrix entries of $R$ and the three body-fixed
angular velocity components of $\Omega$ \cite{Kirchoff}. Since the allowed paths in the enlarged 
space must correspond to paths in the original configuration space $SO(3)$,
they  must satisfy the nine 
constraint equations
\beqa
0 & = & R^t R - I \nonumber \\
0 & = & \dot{R} R^t - \Omega \label{so3constraint}.
\eeqa  
Introduce nine Lagrange multipliers in the form of real $3\times 3$
symmetric $\sigma$ and antisymmetric $\mu$ matrices. Then the Lagrangian on
the twelve dimensional configuration space is given by
\beq
L[R,\dot{R},\Omega,\dot{\Omega}] =  
 T(\Omega)  
 +\hf \mbox{tr}[\sigma\cdot(R^t R-I)]
 +\hf \mbox{tr}[\mu\cdot(\dot{R}R^t-\Omega)].
\eeq
This is just a quadratic function of the 12 coordinates and their
derivatives. To find the equations of motion, note first that the Lagrangian
is independent of $\dot{\Omega}$. Therefore, the Euler-Lagrange equation simplifies to 
$\partial L / \partial \omega_i = 0$, or
\beq
\mu_{ij} = -\varepsilon_{ijk}\frac{\partial T}{\partial \omega_k} . 
\eeq
The derivative of the kinetic energy with respect to the angular velocity
is the angular momentum in the body-fixed frame
\beq
L_k = \omega_k I_k \, .
\eeq
This is an axial vector and is naturally represented by an antisymmetric
matrix $\tilde{L} = \sum L_k \mbox{\fraktur e}_k$; therefore, the Lagrange equation for the ignorable
coordinates says that $\mu$ is the negative of the angular momentum
\beq
\mu = - \tilde{L}.
\eeq
Next, the Lagrange equations for the $R_{ij}$ give the matrix equation
\beq
\dot{\mu}\cdot R + 2\mu\cdot\dot{R} = R\cdot(\sigma+\sigma^t) \label{so3lagrange1}
\eeq
On the constraint hypersurface, the matrix
\beq
\dot{\mu} + 2\mu\cdot \Omega = R\cdot(\sigma+\sigma^t)\cdot R^t
\eeq
is symmetric, or
\beq
\dot{\mu} = [\Omega, \mu] \, . \label{so3lagrange2}
\eeq
Therefore, the dynamical equations for a rigid body are a finite dimensional Lax system \cite{Lax, moser, Fla74}
\beq
\frac{d}{dt}\tilde{L} = [\Omega, \tilde{L}] . 
\eeq
Expressed in vector form, this becomes the usual Euler equation for the
precession of the angular momentum vector
\beq
\frac{d\mathbf{L}}{dt} = - \mathbf{\bbf{\omega}}\times\mathbf{L} . \label{rigidrotation}
\eeq

In addition to rotational symmetry, there is a symmetry of the rigid body due to invariance with respect to certain rotations in the body-fixed frame. While rotational symmetry follows from the invariance of the metric with respect to right multiplication, a body-fixed symmetry corresponds to metric invariance with respect to left multiplication. For $g\in SO(3)$,  left multiplication by $g$ of the metric yields
\beqa
{\textbf g}_{gR} ((L_g)_* (\mbox{\bigscript R}_{\Omega_1})_{\! R}, (L_g)_* (\mbox{\bigscript R}_{\Omega_2})_{\! R}) & = &{\textbf g}_{gR} ( (\mbox{\bigscript R}_{\mbox{\tiny Ad}_g \Omega_1})_{gR}, (\mbox{\bigscript R}_{\mbox{\tiny Ad}_g \Omega_2})_{gR} ) \nonumber \\
& = & - \mbox{tr}\left( Ad_g\Omega_1\ A^2\ Ad_g\Omega_2\right) \nonumber \\
& = & -\mbox{tr}\left( \Omega_1\ (g^{-1} A^2 g)\ \Omega_2 \right) .
\eeqa
When $g$ and $A^2$ commute, $A^2 g = g A^2$, the metric is invariant with respect to left multiplication by $g$. Let $H$ denote the subgroup of $SO(3)$ consisting of the rotations $g$ that commute with $A^2$. If a curve $t\longmapsto R(t)$ minimizes the action, then, for each $g\in H$, so must the curve $t\longmapsto gR(t)$ that has the same action. Hence, the dynamical system factors through the projection onto the right coset space $SO(3)/H$, and the solutions are curves $t\longmapsto H R(t)$ in this coset space. Since the metric is left $H$-invariant on $SO(3)$, it induces a well-defined metric on the right coset space $SO(3)/H$.

There are three possibilities for the subgroup $H$. For a triaxial rotor, the generic situation, the three principal moments are different, say $a_1 > a_2 > a_3 > 0$, and the subgroup $H$ is the dihedral group
\beq
D_2 = \{ m=\mbox{diag}(m_1,m_2,m_3)\ |\ m_i = \pm 1, \det(m)=1 \} .
\eeq
It is a discrete group with four elements corresponding to rotations by $\pi$ about each principal axis. When the rotor is axially symmetric, two moments are equal. For prolate ($a_1 > a_2=a_3 >0$) and oblate ($a_1=a_2>a_3>0$) shapes, the subgroup $H=O(2)$, and it consists of all continuous rotations about the symmetry axis plus a rotation by $\pi$ about a perpendicular nonsymmetry principal axis. In the case of a sphere ($a_1=a_2=a_3$), $H=SO(3)$ and the coset space is a fixed point.

The coset space is geometrically interpreted as the space of orientations of an ellipsoidal body. Consider a rigidly rotating triaxial ellipsoid with fixed axis lengths 
$a_1 > a_2 > a_3 > 0$. Let ${\cal O}$ denote the closed submanifold of ${\cal Q}$ that corresponds to the orientations of this ellipsoid,
\beq
{\cal O} = \{ q=R^t A^2 R \in Q, R\in SO(3) \} \, .
\eeq 
Although each rotation $R$ defines a point of ${\cal O}$, the correspondence
is not one-to-one. If $R_1$ and $R_2$ define the same oriented ellipsoid, 
$q=R_1^t A^2 R_1 = R_2^t A^2 R_2$, then $R_2 R_1^t$ commutes with the
diagonal matrix $A^2$, i.e., $R_2 R_1^t \in D_2$ and these two rotation group elements are in the same
right coset, $D_2 R_2 = D_2 R_1$. Thus the space ${\cal O}$ of orientations
of a triaxial ellipsoid is canonically diffeomorphic to the right coset space 
$SO(3)/D_2$. Moreover, a rigid rotation of the ellipsoid,
$q \longmapsto  r^t q r$ for $r\in SO(3)$, is equivalent to
right multiplication
in the right coset space, $D_2 R \longmapsto D_2 R r$.
If the ellipsoid is axially symmetric, then the configuration manifold is the right coset space $SO(3)/O(2)$. For a sphere, the configuration space is a single point.

\subsection{Vibrations}
If the three axis lengths vibrate, then the number of degrees of freedom must be increased at least by three. To model rotational and vibrational motion, a minimal extension of the theory is attained by requiring that there exists a curve $t\longmapsto R(t)$ in the rotation group and a curve $t\longmapsto A(t)$ in the space ${\cal A}$ of positive-definite diagonal matrices for which $R(t) \mathbf{X}_\alpha (t) = \mathbf{x}_{\alpha}(t)$ depends on time $t$, but $A(t)^{-1} \mathbf{x}_{\alpha}(t) = \mathbf{y}_{\alpha}$ is independent of time. Since all the particles move together according to the same linear transformations $R(t)$ and $A(t)$, the motion is a collective rotation-vibration. As for rigid rotation, the orthogonal matrix $R(t)$ is chosen so that $q(x(t)) = A(t)^2$ is a diagonal matrix. The diagonal matrix $A(t)$ makes the reference  distribution's quadrupole-monopole tensor equal to the identity matrix $q(y) = I$. For a uniform density ellipsoid, the combined rotation $R(t)$ and stretch $A(t)$ transform it into the unit sphere, in units of $R_0$.

The velocity vector for each particle in the inertial frame is $\mathbf{U}_\alpha = (\dot{R}^t A + R^t \dot{A}) \mathbf{y}_\alpha$. The kinetic energy of the rotating and vibrating system is
\beqa
T & = & \sum_{\alpha}\frac{m_\alpha}{2} \mathbf{U}_\alpha \cdot
\mathbf{U}_\alpha \nonumber \\
& = &  (MR_0^2/10)\ (- \mbox{tr} (A^2 \Omega^2) + \mbox{tr}(\dot{A}^2)) \, . \label{kineticrv}
\eeqa

The velocity $V(t)$ is the tangent to the curve $t\longmapsto (R(t), A(t))$ in the direct product space $SO(3)\times{\cal A}$
\beq
V(t) = -\mbox{\bigscript R}_{\Omega(t)} + \sum_k \dot{a}_k(t) \frac{\partial}{\partial a_k} , \label{basevf}
\eeq
where $A = \mbox{diag}(a_1, a_2, a_3)$ and $\Omega = \dot{R} R^t$.
The metric on the rotation group is extended to the direct product manifold so that
$\partial/\partial a_k$ is an orthonormal basis for the tangent spaces of
${\cal A}$:
\beqa
{\textbf g}_{(R,A)} (\frac{\partial}{\partial a_k}, \frac{\partial}{\partial a_l}) & = & \delta_{kl} \nonumber \\
{\textbf g}_{(R,A)}(\mbox{\bigscript R}_{\Omega_1},\mbox{\bigscript R}_{\Omega_2}) & = & - {\rm tr}(\Omega_1 A^2 \Omega_2) \label{productmetric} \\
{\textbf g}_{(R,A)}(\frac{\partial}{\partial a_k}, \mbox{\bigscript R}_{\Omega}) & = & 0 . \nonumber
\eeqa
Hence the kinetic energy Eq.(\ref{kineticrv}) is proportional to the squared length of the velocity
\beq
T  =  (MR_0^2/10)\ {\textbf g}_{(R(t),A(t))} (V(t), V(t)) .
\eeq

The Lagrangian is the difference between the kinetic and potential energies.
The potential energy V(A) is a pure function of the axis lengths. The
Euler-Lagrange equations may be obtained using a similar line of reasoning
as in the rigid body problem by enlarging the configuration space from 12 to
15 dimensions to accommodate the axis lengths. The Lagrange equations corresponding to the axis length vibrations are 
\beq
\frac{M R_0^2}{5} \ddot{a}_i = \frac{\partial T}{\partial a_i} 
- \frac{\partial V}{\partial a_i} . \label{axisvibrations}
\eeq 
The Euler equation Eq.(\ref{rigidrotation}) for rigid body motion is unaltered. 

This dynamical system may be transferred from the direct product space $SO(3)\times{\cal A}$ onto the ellipsoidal space ${\cal Q}$. Consider the mapping
\beqa
\phi: SO(3) \times {\cal A} & \longrightarrow & {\cal Q} \nonumber  \\
(R, A) & \longmapsto & q = \phi (R, A) = R^t A^2 R .
\eeqa 
Since any positive-definite symmetric matrix may be diagonalized by an orthogonal matrix and its eigenvalues are real positive numbers, the mapping $\phi$ is onto ${\cal Q}$. But $\phi$ is many-to-one for two reasons. First, although the positive eigenvalues of $q\in{\cal Q}$ are unique, their ordering is not. For a fixed $A\in{\cal A}$, let $K$ denote the group of all $k\in SO(3)$ such that $k A k^t \in {\cal A}$. Then the transformation $A\longmapsto k A k^t$ permutes the entries of $A=\mbox{diag}(a_1, a_2, a_3)$. If the entries of $A$ are distinct, then $K$ is the discrete subgroup of $SO(3)$ with $24$ elements consisting of the unit determinant matrices with exactly one nonzero entry, equal to $\pm 1$, in each row and in each column. Second, although the eigenspaces of $q$ are unique, the eigenvectors (the rows of $R$) are not. In the generic case of distinct eigenvalues the rows are determined up to a factor of $\pm 1$ and permutations; left multiplication of $R$ by $k\in K$ corresponds to this indeterminacy. Define an action of the group $K$ on the product space by $(R,A)\longmapsto (k R, k A k^t)$ for $k\in K$. It is evident that $q = \phi (R, A) = \phi (k R, k A k^t)$ for all $k\in K$.

The metric at $(R,A)$ is invariant under the transformation group $K$. With respect to the $K$-action, the vector fields $\partial / \partial a_i$ are permuted among themselves, and the right invariant vector field $\mbox{\bigscript R}_\Omega$ is transformed to $\mbox{\bigscript R}_{\mbox{\tiny Ad}_k \Omega}$. Hence the metric of Eq.(\ref{productmetric}) is $K$-invariant. The $K$-invariance of the metric implies that the Riemannian structure is well-defined on the manifold ${\cal Q}$. The Lagrangian is also well-defined on the ellipsoidal space because the potential energy function is assumed invariant under permutations of its arguments, $V(A) = V(k A k^t)$ for $k\in K$.

The model degrees of freedom corresponding to rotations and vibrations are coupled to each other in a simple way. As the body axis lengths vary, the moments of inertia change accordingly and, thereby, the rotation of the body is modified. But, at each instant in
time, the moment of inertia is the rigid body value. For fluids, say
a water droplet, the description must be enhanced so that the moment of inertia
can assume the irrotational flow value. Physically this requires the inclusion
of additional degrees of freedom corresponding to the internal vortex motion
of the fluid. The Riemann ellipsoid model achieves that goal.

\section{RIEMANN ELLIPSOIDS}

The nine-dimensional configuration space of the Riemann ellipsoid model is the connected component of the general linear group, ${\cal {\cal P}} = GL_+(3,\mbox{\fields R})$. The space ${\cal P}$ will be shown to be a principal bundle over the base manifold ${\cal Q}$ with structure group $G$. Locally the bundle ${\cal P}$ is the Cartesian product of ${\cal Q}$ and $G$, and points $\xi$ in ${\cal P}$ are defined by unique pairs $\xi = (q; S)$ for $q\in {\cal Q}$, $S\in G$. As with the rigid rotor problem, the use of moving frames simplifies the geometrical analysis by allowing for tangent space bases  that include the angular momentum and Kelvin circulation vectors. ${\cal P}$ is also a Riemannian manifold whose metric is inherited from the kinetic energy on Euclidean space. The conservation laws for the angular momentum and Kelvin circulation are consequences of the metric's invariance with respect to left multiplication by the rotation group $SO(3)$ and right multiplication by the structure group $G$, respectively. The dynamical equations for Riemann ellipsoids are derived from a Lagrangian that respects the metric's invariance properties.

\subsection{Bundle Structure}
The mapping
$\pi : {\cal P} \longrightarrow {\cal Q}$\, , $\pi(\xi) = \xi \xi^t$, is onto the space of all
ellipsoidal shapes because $\phi$ from $SO(3) \times {\cal A}$ is onto ${\cal Q}$ and $q = \phi (R,A) = R^t A^2 R = \pi (\xi)$ for $\xi = R^t A \in {\cal P}$.

$\pi$ is not one-to-one since ${\cal P}$ is nine dimensional and the ellipsoidal space ${\cal Q}$ is six dimensional.
If both $\xi_1$ and $\xi_2$ define the same ellipsoid, then 
$\pi(\xi_1)=\pi(\xi_2)$ and $(\xi_2^{-1}\xi_1) (\xi_2^{-1}\xi_1)^t = I$,
or $\xi_2^{-1}\xi_1$ is an orthogonal matrix. Thus $\xi_1$ and $\xi_2$
define the same ellipsoid if and only if $\xi_1$ and $\xi_2$ are in the
same left coset of $G=SO(3)$, i.e., $\xi_1 \in \xi_2 G$. There is a one-to-one
and onto identification of the space of ellipsoidal surfaces with the left coset space
\beq
{\cal Q} \cong {\cal P}/G = GL_+(3,\mbox{\fields R})/SO(3) \, .
\eeq 
                             
By choosing a smooth set of left coset representatives, the space ${\cal P}$ may be identified locally 
with the Cartesian product of ${\cal Q}$ and $G$. Such a set is provided by
a smooth mapping $\tau$ from an open neighborhood ${\cal U}_\tau\subset{\cal Q}$ into ${\cal P}$ such that $\tau(q_1) G = \tau(q_2) G$ if and only if $q_1=q_2$, for $q_1, q_2 \in {\cal U}_\tau$.  
Given a smooth coset representative mapping, a diffeomorphism between ${\cal U}_\tau\times G$ and an open neighborhood ${\cal V}_\tau$ of ${\cal P}$ is defined by
\beqa
{\cal U}_\tau\times G \subset {\cal Q}\times G & \longrightarrow & {\cal V}_\tau \subset {\cal P} \nonumber \\
(q; S) & \longmapsto & \xi = \tau(q) S  \label{localchart}
\eeqa
for $q\in {\cal U}_\tau$ and $S\in G$.
The identification is not canonical since it depends on the arbitrary choice of the coset
representative map $\tau$. The smooth choice of coset representatives $\tau(q) = R^t A$ for $q = R^t A^2 R$ is a well-defined one-to-one mapping only locally because the diagonalization of $q$ is not unique. There is no {\em smooth} mapping $\tau$ defined globally on ${\cal Q}$. The space ${\cal P}$ is only locally diffeomorphic to the Cartesian product of ${\cal Q}$ and $G$. Throughout this paper, whenever a chart for ${\cal P}$ is needed, a smooth choice of coset representatives of the form  $\tau(q) = R^t A$ is taken in some open neighborhood ${\cal U}_\tau$ of ${\cal Q}$, and the points of the bundle are given in an open neighborhood ${\cal V}_\tau$ of ${\cal P}$ by $\xi = R^t A S$. This choice has the advantage of distinguishing among rotational, vibrational, and vortex motions. 

Consider the right multiplication of $G$ on ${\cal P}$:
$R_g(\xi) = \xi g^{-1}$ for $\xi\in {\cal P}$ and $g\in G$. $G$ must be
distinguished from the rotation group that
acts by left multiplication on ${\cal P}$. The point $q\in {\cal Q}$ corresponding to both $\xi$ and 
$R_g(\xi)$ is the same. Thus the transformation $R_g$ does not change the
ellipsoidal boundary of the fluid, but it does change its internal motion. In terms of the local chart of Eq.(\ref{localchart}) for ${\cal P}$, the right multiplication is $R_g(q; S) = (q; Sg^{-1})$.

The space ${\cal P}$ is a principal fiber bundle over the base manifold ${\cal Q}$ with structure group
$G$, since the manifold ${\cal P}={\cal P}({\cal Q}, G, \pi)$ satisfies the following
defining properties of a principal $G$-bundle \cite{KN}:
\begin{itemize}
\item The space ${\cal P}$ is locally diffeomorphic to the Cartesian product of
the base manifold and the structure group. Given a coset representative
map, a bundle point in the corresponding local chart is $\xi = (q; S)$, where $q\in{\cal Q}$ and  
$S\in G$. Such charts are called {\em local trivializations}.
\item The projection $\pi : {\cal P}\longrightarrow {\cal Q}$ is defined from the
bundle onto the base manifold: $\pi(\xi) = \xi \xi^t = q\in{\cal Q}$ for $\xi=(q; S)\in {\cal P}$.
The inverse image of a fixed point $q\in{\cal Q}$ in the base manifold is 
diffeomorphic to the structure group, $\pi^{-1}(q) \cong G$, which is called the {\em fiber} over $q$. 
\item A right multiplication $R_g:{\cal P}\longrightarrow {\cal P}$ is given by
$R_g(\xi) = \xi g^{-1}$ and it satisfies $\pi \circ R_g = \pi$. In terms of the local chart, $R_g(q; S) = (q; S g^{-1})$ for $\xi=(q; S)\in {\cal P}$. 
\end{itemize} 

\subsection{Tangent Vector}

Consider a curve $t\longmapsto \xi(t)$ in the bundle ${\cal P}$. Such a curve may be 
identified with the collective motion of a many-body system for which the trajectory of each particle is constrained by $\mathbf{X}_\alpha (t) = \xi(t) \mathbf{y}_{\alpha}$, where $\mathbf{y}_{\alpha}$ is independent of time. 
The reference distribution $\mathbf{y}_\alpha$ is chosen so that its
quadrupole-monopole tensor is the identity
matrix, $q(y) = I$. 
With this choice the instantaneous dimensionless quadrupole-monopole tensor
simplifies to $q(X) = \xi \xi^t$. Hence the base manifold point $\pi(\xi)$ is interpreted physically as the quadrupole-monopole tensor of the constrained many-body system.

The velocity vector for each particle is $\mathbf{U}_\alpha = \dot{\xi}
\mathbf{y}_\alpha = \dot{\xi} \xi^{-1} \mathbf{X}_\alpha$. Note that the velocity of each particle is a linear function of its position vector. The velocity vector may be expressed as the value of a right invariant vector field on the group ${\cal P}$ at the point $\xi$,
\beqa
V(t) & = & \sum_{ij} \dot{\xi}_{ij} \frac{\partial}{\partial \xi_{ij}} 
\nonumber \\
& = & \sum_{ij} (\dot{\xi}\xi^{-1}\cdot \xi)_{ij} \frac{\partial}{\partial \xi_{ij}} \nonumber \\
& = & -(\mbox{\bigscript R}_u)_\xi, \mbox{\ for\ }u = \dot{\xi}\xi^{-1} .
\eeqa
The vector field $\mbox{\bigscript R}_u$ is right invariant with respect to the entire
group ${\cal P}$, in contrast to the invariant vector fields introduced in $\S 2$ for
the rotation group. Although nominally derived for the rotation group, the results of $\S 2$ hold for any Lie group. This includes the concepts of left and right invariant vector fields and their transformation properties under left and right translation. The velocity $u$ is an element of the Lie algebra $M_3(\mbox{\fields R})$
of $GL_+(3,\mbox{\fields R})$, i.e., a real $3\times 3$ matrix.

A curve $t\longmapsto (R(t), A(t), S(t))$ in the direct product manifold $SO(3) \times {\cal A} \times G$ defines a curve $t\longmapsto \xi(t) = R(t)^t A(t) S(t)$ in ${\cal P}$. The curve $t\longmapsto S(t)$ in the structure group $G$ defines the vorticity $\Lambda = \dot{S}S^{-1}$. $\Lambda$ is
an element of the Lie algebra {\fraktur g} of the group $G$ and must be an antisymmetric
matrix. The curve $t\longmapsto q(t) = \pi(\xi(t)) = R(t)^t A(t)^2 R(t)$ is in the base manifold ${\cal Q}$ and describes the rotation and vibration of an ellipsoid.

However, any segment of a smooth curve $t\longmapsto \xi(t)$ in the bundle ${\cal P}$ cannot be lifted uniquely to a smooth curve in the direct product manifold. Recall that the ordering of the elements of $A$ is not unique. As discussed in $\S 2$, the ambiguity in the lifting is described by the subgroup $K$ of the rotation group: for every $k\in K$, $\xi = R^t A S$ is invariant with respect to the transformation $(R, A, S) \longmapsto (kR, kAk^t, kS)$ in the direct product space.

The velocity $V(t)$ may be simplified to the sum of terms corresponding to rotational, vibrational, and vortex motions:
\beq
u = \dot{\xi} \xi^{-1} = R^t \left( -\Omega + A^{-1}\dot{A} 
+ A\Lambda A^{-1} \right) R . \label{threeterms}
\eeq
Each of these three vector fields on the bundle ${\cal P}$ may be expressed in the $SO(3) \times {\cal A} \times G$ chart ($\xi = R^t A S$) as
\beqa
(\mbox{\bigscript R}_{R^t \Omega R})_\xi & = & -(\mbox{\bigscript R}_{\Omega})_{\! R} =  (\Omega R)_{ij}\left(\frac{\partial}{\partial R_{ij}}\right)_{\! R} \mathletter{a} \label{rotterm}\\
(\mbox{\bigscript R}_{R^t A^{-1}\dot{A} R})_\xi & = & -\dot{a}_i \left(\frac{\partial}{\partial a_i}\right)_A  \mathletter{b}\label{vibterm} \\
(\mbox{\bigscript R}_{R^t A\Lambda A^{-1} R})_\xi & = & (\mbox{\bigscript R}_{\Lambda})_S = (\Lambda S)_{ij}\left(\frac{\partial}{\partial S_{ij}}\right)_{\! S}, \label{vorterm}  \mathletter{c} 
\eeqa
and the velocity simplifies to
\beq
V(t)  =  -(\mbox{\bigscript R}_{\Omega})_{\! R} +\dot{a}_i \left(\frac{\partial}{\partial a_i}\right)_A - (\mbox{\bigscript R}_{\Lambda})_S \, . \label{gl3velocity}
\eeq 
Note that $(\mbox{\bigscript R}_\Omega)_{\! R}$ is a right invariant vector field on $SO(3)$ and
$(\mbox{\bigscript R}_\Lambda)_S$ is a right invariant vector field on $G$. This common notation
for all right invariant vector fields should not be confusing since $\Omega$
is reserved for the Lie algebra $\mbox{\fraktur so}(3)$ of $SO(3)$ while $\Lambda$ is reserved for the Lie algebra {\fraktur g} of the vortex group $G$. A subscript $\xi$ indicates an invariant vector field on ${\cal P}$ evaluated at $\xi$. The subscript $R$ (respectively, $S$) indicates an invariant vector field on $SO(3)$ evaluated at $R$ (respectively, on $G$ evaluated at $S$).

To prove Eq.(\ref{gl3velocity}), consider each of the three terms of $u$ in Eq.(\ref{threeterms}) separately.
For the rotational contribution to $V(t)$, Eq.(\ref{rotterm}),
\beqa
(\mbox{\bigscript R}_{R^t \Omega R})_\xi & = & ((R^t \Omega R) \xi)_{ij}\frac{\partial}{\partial \xi_{ij}} \nonumber \\
& = & (R^t \Omega A S)_{ij} \frac{\partial}{\partial \xi_{ij}} \nonumber \\
& = & (R^t \Omega)_{ik} (A S)_{kj} \frac{\partial}{\partial \xi_{ij}} \nonumber \\
& = & (\Omega R)_{ki} \frac{\partial}{\partial R_{ki}} \nonumber \\
& = & -(\mbox{\bigscript R}_\Omega)_{ R} ,
\eeqa
because, according to the chain rule, 
\beqa
\frac{\partial}{\partial R_{ij}} & = & \frac{\partial \xi_{kl}}{\partial R_{ij}} 
\frac{\partial}{\partial \xi_{kl}} \nonumber \\
& = & \delta_{mi} \delta_{kj} (AS)_{ml} \frac{\partial}{\partial \xi_{kl}} \nonumber \\
& = & (A S)_{il} \frac{\partial}{\partial \xi_{jl}} \, .
\eeqa
Similar arguments apply to the vibrational and vortex terms, Eqs.(\ref{vibterm},\ref{vorterm}).

\subsection{Riemannian Structure}
The kinetic energy of the linear velocity field is
\beqa
T & = & \sum_{\alpha}\frac{m_\alpha}{2} \mathbf{U}_\alpha \cdot
\mathbf{U}_\alpha \nonumber \\
& = & (MR_0^2/10)\, \mbox{tr}(\dot{\xi} \dot{\xi}^t)  \, .
\eeqa

For $X$, $Y$ in $M_3(\mbox{\fields R})$, define the metric at the point $\xi \in {\cal P}$ by
\beq
{\textbf g}_\xi((\mbox{\bigscript R}_X)_\xi, (\mbox{\bigscript R}_Y)_\xi) = \mbox{tr} (X (\xi \xi^t) Y^t) .
\eeq
This is a positive-definite bilinear form defined on each tangent space of ${\cal P}$; hence, 
${\cal P}$ is a Riemannian manifold. The kinetic energy is given in terms of the
metric by
\beq
T = (MR_0^2/10)\, {\textbf g}_\xi (V(t), V(t)) .
\eeq
Expanding the velocity into the three types of motion, Eq.(\ref{gl3velocity}), simplifies the kinetic energy to
\beq
T = (MR_0^2/10) \left(-\mbox{tr}(A^2\Omega^2) + \mbox{tr} (\dot{A}^2)
- \mbox{tr}(A^2 \Lambda^2) + 2\, \mbox{tr}(\Omega A \Lambda A) \right) . \label{Riemannkinetic}
\eeq
Note that the last term is a Coriolis coupling between the rotational
and vortex degrees of freedom.

\subsection{Lagrangian}
Riemann ellipsoid dynamics may be derived using the Lagrange formalism. The potential energy $V$ is assumed to be a rotational scalar function on the base manifold, i.e., a smooth function $V(A)$ of the axis lengths that is invariant under $K$, $V(k A k^t) = V(A)$ for all $k\in K$. The rotation, vibration, and vortex motions of a Riemann ellipsoid correspond to the path in ${\cal P}$ that minimizes the action with respect to all smooth curves connecting two fixed points.

The Lagrangian is invariant with respect to left multiplication by the rotation group $SO(3)$ and with respect to right multiplication by the structure group $G$. The potential energy is obviously invariant with respect to both of these multiplications. The kinetic energy is invariant because the metric is invariant: if $r\in SO(3)$, then left multiplication of the metric yields
\beqa
{\textbf g}_{r\xi} ( (L_r)_* \mbox{\bigscript R}_X, (L_r)_* \mbox{\bigscript R}_Y ) & = & {\textbf g}_{r\xi} ( \mbox{\bigscript R}_{\mbox{\tiny Ad}_r X}, \mbox{\bigscript R}_{\mbox{\tiny Ad}_r Y} ) \nonumber \\
& = & \mbox{tr}\left( (rXr^{-1})  r\xi (r\xi)^t (rYr^{-1})^t \right) \nonumber \\
& = & {\textbf g}_\xi(\mbox{\bigscript R}_X, \mbox{\bigscript R}_Y) .
\eeqa
If $g\in G$, then right multiplication of the metric gives
\beqa
{\textbf g}_{\xi g^{-1}} ( (R_g)_* \mbox{\bigscript R}_X, (R_g)_* \mbox{\bigscript R}_Y ) & = & {\textbf g}_{\xi g^{-1}} ( \mbox{\bigscript R}_X, \mbox{\bigscript R}_Y ) \nonumber \\
& = & \mbox{tr}\left( X  (\xi g^{-1}) (\xi g^{-1})^t Y^t \right) \nonumber \\
& = & {\textbf g}_\xi (\mbox{\bigscript R}_X, \mbox{\bigscript R}_Y) . 
\eeqa
By Noether's theorem, the invariances imply conservation laws. Left invariance implies conservation of the angular momentum vector $\mathbf{L}$, while right invariance requires conservation of the Kelvin circulation vector $\mathbf{C}$. The Lagrangian is also invariant under time translation which implies conservation of energy $E = T + V$.

As with the rigid body problem, it is preferable for computational purposes to enlarge the configuration space ${\cal P}$ to the $27$-dimensional Euclidean space whose coordinates are the 18 matrix entries of $R$ and $S$, the six  angular and vortex velocity components of $\Omega$ and $\Lambda$, and the three entries of the diagonal matrix $A$ \cite{Shieh}. This extension avoids the introduction of Euler angles for $SO(3)$ and $G$, Moreover, two conservation laws follow because the Lagrangian does not depend explicitly on the angular and vortex velocities. Since the allowed paths in the enlarged space must correspond to paths in the original configuration space ${\cal P}$,
they must satisfy eighteen constraint equations. In addition to the $SO(3)$ constraints of Eq.(\ref{so3constraint}), there are also constraints for the structure group:
\beqa
0 & = & S^t S - I \nonumber \\
0 & = & \dot{S} S^t - \Lambda \label{gconstraint}.
\eeqa  
Eighteen Lagrange multipliers must be introduced in the form of real $3\times 3$
symmetric matrices $\sigma$ and $\tau$, and antisymmetric matrices $\mu$ and $\nu$. The Lagrangian $L = L[R,\dot{R},\Omega,\dot{\Omega}, A, \dot{A}, S,\dot{S},\Lambda, \dot{\Lambda}]$ on the twenty-seven dimensional configuration space is given by
\beqa
L & = & 
 T( \Omega, A, \dot{A}, \Lambda )  - V(A) \nonumber \\
 & & +\hf \, \mbox{tr}[\sigma\cdot(R^t R-I)] + \hf \mbox{tr}[\tau\cdot(S^t S-I)] \nonumber \\
 & & +\hf \mbox{tr}[\mu\cdot[\dot{R}R^t-\Omega)] + \hf \mbox{tr}[\nu\cdot[\dot{S}S^t-\Lambda) ]. 
\eeqa
Except for the potential energy, the Lagrangian is just a quadratic function of the 27 coordinates and their
derivatives.

The conservation laws are derived by noting that $L$ is independent of $\dot{\Omega}$ and $\dot{\Lambda}$, and, therefore, the Euler-Lagrange equations for the angular and vortex velocities simplify to 
\beq
\mu_{ij} =  -\varepsilon_{ijk}\frac{\partial T}{\partial \omega_k}  \mbox{\ \ and\ \ } \nu_{ij}  =  -\varepsilon_{ijk}\frac{\partial T}{\partial \lambda_k} . 
\eeq
The derivatives of the kinetic energy with respect to the angular velocity and vortex velocity are the vectors of angular momentum and circulation, respectively:
\beqa
L_k & = &\  \ \frac{\partial T}{\partial \omega_k} =  (M R_0^2/5)\, [ (a_i^2+a_j^2)\omega_k -2a_ia_j\lambda_k ]  \mathletter{a} \\
C_k & = & -\frac{\partial T}{\partial \lambda_k} = (M R_0^2/5)\, [ 2a_ia_j\omega_k  - (a_i^2+a_j^2)\lambda_k ], \mathletter{b}
\eeqa
where $i,j,k$ are cyclic. Note that the angular momentum depends on both the angular and vortex velocities. Because the two vectors are represented naturally by antisymmetric
matrices, $\tilde{L} = \sum L_k \mbox{\fraktur e}_k$ and $\tilde{C} = \sum C_k \mbox{\fraktur e}_k$, Lagrange's equations reduce to 
\beq
\mu = - \tilde{L} \mbox{\ \ and\ \ } \nu =  \tilde{C}.
\eeq
Lagrange's equations for the $R_{ij}$ and $S_{ij}$ are derived using arguments similar to Eqs.(\ref{so3lagrange1} - \ref{so3lagrange2}) and give the matrix equations
\beqa
\dot{\mu} = [\Omega, \mu]  \mbox{\ \ and\ \ } \dot{\nu} = [\Lambda, \nu] \, .
\eeqa
Hence, the dynamical equations for the rotational and vortex motion of a Riemann ellipsoid body are a Lax system
\beq
\frac{d}{dt}\tilde{L} = [\Omega, \tilde{L}]   \mbox{\ \ and\ \ } \frac{d}{dt}\tilde{C} = [\Lambda, \tilde{C}].
\eeq
Expressed in vector form, the precessions of the angular momentum vector and the Kelvin  circulation vector are 
\beq
\frac{d\mathbf{L}}{dt} = - \mathbf{\bbf{\omega}}\times\mathbf{L}   \mbox{\ \ and\ \ }  \frac{d\mathbf{C}}{dt} = - \mathbf{\bbf{\lambda}}\times\mathbf{C} . \label{riemannrotation}
\eeq
The two vector conservation laws are
\beq
\frac{d}{dt}\left( R^t \mathbf{L} \right) = 0  \mbox{\ \ and\ \ } \frac{d}{dt}\left( S^t \mathbf{C} \right) = 0 . 
\eeq

For the axis lengths vibrations, Lagrange's equations in the variables $a_k$ yield the same formal equation (\ref{axisvibrations}), but the Riemann kinetic energy is substituted for the rigid rotor expression. The entire dynamical system for Riemann ellipsoids may be expressed as a finite-dimensional Lax system \cite{ROS93b}.

\section{CONNECTIONS ON THE RIEMANN ELLIPSOID BUNDLE}
For many interacting systems, there are constraint forces in addition to those described by the potential energy $V(A)$. The simplest case is the rigid body for which the vortex velocity vanishes $\mathbf{\bbf{\lambda}}=0$. This holonomic constraint simplifies the dynamical system to the rigid body theory of $\S 2$. But constraints are not typically holonomic. For example, an irrotational fluid (like a water droplet) has zero circulation, $\mathbf{C}=0$. Another example is the falling cat \cite{Kane}, for which the angular momentum vanishes, $\mathbf{L} = 0$. In these cases the vortex velocity is proportional to the angular velocity \cite{Chandrasekhar69}
\beqa
\mbox{\ irrotational flow: \ }\lambda_k & = & \frac{2a_ia_j}{a_i^2+a_j^2}\omega_k,   \mathletter{a} \\ 
\mbox{\ falling cat: \ }\lambda_k & = & \frac{a_i^2+a_j^2}{2a_ia_j} \omega_k,  \mathletter{b} 
\eeqa
where $i, j, k$ are cyclic. For a holonomic constraint the structure group dynamical equation, $\dot{S} = \Lambda S$, may be solved explicitly for $S(t)$. For example, if $\Lambda = \sum \lambda_k \mbox{\fraktur e}_k$ is constant, then $S(t) = \exp(t \Lambda) S(0)$, where $S(0)$ denotes the value of the structure group element at $t=0$. Unless the axis lengths and angular velocity vector are constant in time, the irrotational flow and falling cat problems are not generally integrable systems. The typical nonholonomic constraint for a Riemann ellipsoid is a proportionality between the vortex and angular velocity components, $\lambda_k = f_k(A) \omega_k$ with a factor $f_k(A)$ that depends on the axis lengths.

Suppose a nonholonomic constraint determines the vortex velocity in terms of the axis lengths and the angular velocity vector. A curve $\gamma : t\longmapsto q(t)$ in the base manifold ${\cal Q}$ defines curves $\tilde{\gamma} : t\longmapsto \xi(t) = (q(t); S(t))$ in the bundle ${\cal P}$ known as {\em horizontal lifts}. To obtain the horizontal lift, take the unique tangent vector field, Eq.(\ref{basevf}), of the curve $\gamma$ in the base manifold, and express the vortex velocity $\Lambda(t)$ in terms of the angular velocity $\Omega(t)$ and vibrational velocity $\dot{A}(t)$ by applying the nonholonomic constraint. Eq.(\ref{gl3velocity}) defines a vector field on the bundle that integrates to the horizontal lift $\tilde{\gamma}$ through the initial point $S(0)\in G$. The bundle vector field is only defined on the subbundle $\pi^{-1}(q(t))$ lying above the base manifold curve. Note that the horizontal lift depends on the association of a unique vector field on the bundle to a vector field on the base manifold. As it will be seen later, this association is the differential structure, or connection, on a bundle.

The concept of a horizontal lift is physically natural. It says that a many-body system responds to rotations and vibrations (described by a curve $\gamma$ in the base manifold) by internal vortex motions (described by a curve $\tilde{\gamma}$ in the bundle). This response is determined typically by a nonholonomic constraint that depends ultimately on the nature of the forces between the particles. The mathematical description of the response is a horizontal lift $\tilde{\gamma}$ in the bundle, see Figure 3.

The horizontal lift is determined by a connection in the principal bundle. Since the theory of connections is the foundation of differential geometry, a brief review will be given here. The theory of Riemann ellipsoids is regarded as a paradigm for the general gauge theory of two interacting sets of physical degrees of freedom, e.g., quadrupole and pairing \cite{Kumar}, symplectic\cite{ROW85} and interacting boson\cite{IBM}, etc. Although the theory of bundle connections is well known in particle physics, it is less familiar in many-body science.

Suppose ${\cal P}={\cal P}({\cal Q}, G, \pi)$ is a principal bundle over the base manifold ${\cal Q}$ with structure group $G$ and right multiplication $R_g (\xi) = \xi \cdot g$ for all $\xi\in {\cal P}$ and $g\in G$. $R_g$ is assumed to be a group homomorphism, $R_{g_{1}} R_{g_{2}} = R_{g_{1} g_{2}}$, or $(\xi \cdot g_1) \cdot g_2 = \xi \cdot (g_1 g_2)$. The Riemann ellipsoid bundle ${\cal P}=GL_+(3, \mbox{\fields R})$ is an example, but the theory of connections will be cast in its general setting. In a local trivialization, a bundle point is $\xi = (q; S)$, where $q\in{\cal Q}$ and  
$S\in G$, the projection map is $\pi (q; S) = q$, and the right multiplication is $R_g(q; S) = (q; S g^{-1})$.

For each point $\xi$ in the bundle, denote the tangent space to ${\cal P}$ by $T_{\xi}{\cal P}$. The {\em vertical} space $V_\xi$ is a subspace of $T_{\xi}{\cal P}$ consisting of the tangents to curves $t\longmapsto (q; S(t))$ in the fiber ($q\in {\cal Q}$ is fixed). It may be defined without reference to a local trivialization by
\beq
V_{\xi} = \left\{ X\in T_{\xi}{\cal P} \ | \ \pi_{\ast} X = 0  \right\} .
\eeq

Among the vertical vectors, the left invariant vector fields on the structure group are singled out for special consideration. Suppose that the structure group is a Lie subgroup of a real matrix group $GL(d,\mbox{\fields R})$, and the Lie algebra {\fraktur g} of the group $G$ may be identified with an algebra of matrices. If $\Lambda \in$ {\fraktur g} is a Lie algebra element, then the fundamental vector field, denoted by $\Lambda^\ast$, is the left invariant vector field  $\mbox{\bigscript L}_{\Lambda}$ on the fiber $G$. If $\{\mbox{\fraktur f}_a, a=1,\dots,p \}$ is a basis for the $p$-dimensional Lie algebra, then $\{(\mbox{\fraktur f}_a^\ast)_S, a=1,\dots,p \}$ is a basis for the tangent space to the structure group at the point $S\in G$. The fundamental vector fields define a Lie algebra homomorphism, $[\Lambda_1,\Lambda_2]^\ast  =  [ \Lambda_1^\ast, \Lambda_2^\ast ]$, for $\Lambda_1, \Lambda_2 \in$ {\fraktur g}. The right multiplication in the bundle satisfies $\pi \circ R_g = \pi$ and, hence, $\pi_\ast \circ (R_g)_\ast = \pi_\ast$. Thus $(R_g)_\ast$ is a linear transformation from the vertical space $V_\xi$ onto the vertical space $V_{\xi\cdot g}$, and, according to Eq.(\ref{adjoint1}), it is given by $(R_g)_\ast (\Lambda^{\ast})_S = ( Ad_{g} \Lambda )^\ast_{Sg^{-1}}$.

A {\em connection} is a smooth assignment $\Gamma$ of a subspace $H_\xi$ of the tangent space $T_\xi {\cal P}$ to each point $\xi\in {\cal P}$. $H_\xi$ is called the {\em horizontal subspace}. A vector $X \in T_\xi {\cal P}$ is called {\em vertical}, respectively {\em horizontal}, if it lies in $V_\xi$, respectively in $H_\xi$. To be a connection, the assignment must satisfy two properties:
\beqa 
(1)\ T_\xi {\cal P} & = & H_\xi \oplus V_\xi \mbox{\ (direct sum)}  \mathletter{a} \\
(2)\ H_{\xi \cdot g} & = & (R_g)_\ast H_\xi \,  .  \mathletter{b} \label{secondprop}
\eeqa
The first condition means that every tangent vector $X\in T_\xi {\cal P}$ can be uniquely written as
\beq
X = Y + Z, \mbox{\ \ where\ }Y\in H_\xi \mbox{\ \ and\ } Z\in V_\xi ,
\eeq
and $Y$, respectively $Z$, are called the {\em vertical}, respectively {\em horizontal}, {\em components} of $X$. The horizontal projection $Y\in H_\xi$ of $X\in T_\xi {\cal P}$ is denoted by $Y = hX$.

Since the kernel of $\pi_\ast$ at $\xi\in {\cal P}$ is the vertical subspace $V_\xi$, the image of $\pi_\ast$ is $T_q {\cal Q}$, and the tangent space is a direct sum of the vertical and horizontal subspaces, the linear transformation $\pi_\ast$ is an isomorphism from the horizontal subspace onto the tangent space of the base manifold $\pi_\ast : H_\xi \longrightarrow T_q{\cal Q}$\,, where $q =\pi(\xi)$. If $T\in T_q{\cal Q}$ is a tangent vector to the base manifold, then its {\em horizontal lift} is the unique horizontal vector $\tilde{T} \in H_\xi$ such that $\pi_\ast {\tilde T} = T$. Given a basis of smooth vector fields to the $n$-dimensional base manifold at $q\in{\cal Q}$, $\{(\mbox{\bf e} _i)_q, i=1,\dots,n \}$, their unique horizontal lifts are denoted by $\{(\mbox{\bf E}_i)_\xi,  i=1,\dots,n \},\mbox{\ where\ } \pi(\xi)=q$. The set $\{(\mbox{\bf E}_i)_\xi \}$ is a basis for the horizontal subspace $H_\xi$ and
\beq
(\mbox{\bf E}_i)_\xi = (\mbox{\bf e} _i)_q - \sum_a \Gamma^a_i(\xi) (\mbox{\fraktur f}_a^\ast)_S ,
\eeq
where, in a local trivialization, $\xi = (q; S)$. The coefficients $\Gamma^a_i$ must be smooth real-valued functions on the bundle ${\cal P}$.

The second defining property of a connection, Eq.(\ref{secondprop}), asserts that $(\mbox{\bf  E}_i)_{\xi \cdot g} = (R_g)_\ast (\mbox{\bf E}_i)_\xi$. In particular, when $\xi = (q; e)$, where $e$ is the group identity and $g = S^{-1} \in G$, the right translation of a horizontal basis vector at the structure group identity, $(\mbox{\bf E}_i)_{(q; S)} = (R_{S^{-1}})_\ast (\mbox{\bf E}_i)_{(q; e)}$, is
\beqa
(\mbox{\bf E}_i)_{(q; S)} & = & (\mbox{\bf e} _i)_q - \sum_a \Gamma^a_i(q; e) (Ad_{S^{-1}}\mbox{\fraktur f}_a)^\ast_S \nonumber \\
& = & (\mbox{\bf e} _i)_q + \sum_a \Gamma^a_i(q) (\mbox{\bigscript R}_{\mbox{\fraktur f}_a})_S ,
\eeqa
where Eqs.(\ref{lefttoright}, \ref{adjoint1}) are used. Hence, the right invariance of the horizontal spaces implies that the functions $\Gamma$ are determined by their values at $\xi=(q; e)$. The functions $\Gamma^a_i(q) \equiv \Gamma^a_i(q; e)$ are called {\em Christoffel symbols}.

Suppose $q(t)$ is a smooth curve in the base manifold ${\cal Q}$ and its tangent is given by $T(t) = \sum T^i(t) (\mbox{\bf e}_i)_{q(t)}$. The tangent to the horizontally lifted curve $t \longmapsto (q(t); S(t))$ is $\tilde{T} = \sum T^i(t) (\mbox{\bf E}_i)_{(q(t); S(t))}$, and it may be decomposed into horizontal and vertical components
\beq
\tilde{T} = T + \sum \dot{S}_{jk} \frac{\partial}{\partial S_{jk}} ,
\eeq
where $S_{jk}$ are the matrix entries in $GL(d, \mbox{\fields R})$. The components $\dot{S}_{jk}$ are given by
\beqa  
\dot{S}_{jk} & = & \tilde{T} (S_{jk}) \nonumber \\
& = & \sum_i T^i(t) \sum_a \Gamma^a_i(q(t)) (\mbox{\bigscript R}_{\mbox{\fraktur f}_a})_{S(t)} (S_{jk}) \nonumber \\
& = & - \sum_i T^i(t) \sum_a \Gamma^a_i(q(t)) (\mbox{\fraktur f}_a S)_{jk} ,
\eeqa
where Eq.(\ref{diffop}) is applied.
In matrix form the differential equation for the horizontally lifted curve is
\beq
\dot{S} = - \left( \sum_i T^i(t) \sum_a \Gamma^a_i(q(t)) \mbox{\fraktur f}_a \right) S .
\eeq
If the term in parentheses is independent of time, then the differential equation integrates to an exponential.

This finishes the required mathematical preliminaries. The theory will be applied now to the Riemann ellipsoid model. Let {\fraktur e}$_i$ denote, as before, the natural basis for the algebra of $3\times 3$ antisymmetric matrices, $(${\fraktur e}$_i)_{jk} = \varepsilon_{ijk}$. A basis $\{(\mbox{\bf e}_i)_q, i=1,\ldots, 6 \}$ for the tangent space at $q\in Q$ consists of the three right invariant vector fields $(\mbox{\bf e}_i)_q = (\mbox{\bigscript R}_{\mbox{\fraktur e}_i})_{\! R}$ on the rotation group $SO(3)$ and the three vibrational vector fields $(\mbox{\bf e}_{i+3})_q = (\partial / \partial a_i)_A$ on ${\cal A}$ for $i=1,2,3$. A tangent vector to a curve in the base manifold is given by Eq.(\ref{basevf}) or
\beq
T = \sum_{i=1}^3 \left(-\omega_i \,(\mbox{\fraktur e}_i)_q + \dot{a}_i \, (\mbox{\fraktur e}_{i+3})_q \right) .
\eeq
A basis for the Lie algebra {\fraktur g} of the structure group is the set $\{ \mbox{\fraktur f}_i = \mbox{\fraktur e}_i, i=1, 2, 3\}$. The horizontal lift of the curve is given by Eq.(\ref{gl3velocity}) or
\beqa
\tilde{T} & = & \sum_{i=1}^3 \left(-\omega_i \,(\mbox{\fraktur e}_i)_q + \dot{a}_i \, (\mbox{\fraktur e}_{i+3})_q  - \lambda_i \, (\mbox{\bigscript R}_{\mbox{\fraktur f}_i})_S \right) \nonumber \\
& = & - \sum_{i=1}^3 \omega_i \,\left((\mbox{\fraktur e}_i)_q   + \frac{\lambda_i}{\omega_i} \, (\mbox{\bigscript R}_{\mbox{\fraktur f}_i})_S \right) + \sum_{i=1}^3 \dot{a}_i \, (\mbox{\fraktur e}_{i+3})_q .
\eeqa
Therefore the tangent vectors
\beqa
\mbox{\bf E}_i & = & (\mbox{\bigscript R}_{\mbox{\fraktur e}_i})_{\! R}   + \left(\frac{\lambda_i}{\omega_i}\right) \, (\mbox{\bigscript R}_{\mbox{\fraktur f}_i})_S  \mathletter{a}  \\
\mbox{\bf E}_{i+3} & = & \frac{\partial}{\partial a_i} , \mathletter{b}
\eeqa
for $i = 1, 2, 3$, are a basis for the horizontal subspace. They show that the Christoffel symbols for Riemann ellipsoids vanish for the vibrational vectors and simplify to a diagonal form for the rotational vectors
\beq
\Gamma^a_i(q) = \delta^a_i \left( \frac{\lambda_i}{\omega_i} \right) .
\eeq
In particular, the special rotational modes correspond to the following Christoffel symbols:
\beq
\Gamma^k_k  = \left\{ \begin{array}{cl}
0 & \mbox{\ rigid}  \\
2a_ia_j/(a_i^2+a_j^2) & \mbox{\ irrotational}  \\ 
(a_i^2+a_j^2)/(2a_ia_j) & \mbox{\ falling cat},
\end{array} \right. 
\eeq
where $i, j, k$ are cyclic. Note that the irrotational Christoffel symbols are at most one and the falling cat's are at least one.

\subsection{Riemannian Connection}
The irrotational flow and falling cat connections arise naturally from the Riemannian structure and the group structure, respectively. The horizontal subspace $H_{\xi}$ for irrotational flow is defined as the orthogonal complement to the vertical subspace $V_{\xi}$. Denote the vector space of all $3\times 3$ symmetric matrices by {\fraktur m}. The orthogonal complement $V_\xi^\perp$ is given explicitly by
\beq
H_{\xi} = \left\{ (\mbox{\bigscript R}_Y)_{\xi} \in T_{\xi}{\cal P} \ | \ Y\in \mbox{\fraktur m} \right\} .
\eeq
To prove this, suppose $(\mbox{\bigscript R}_\Lambda)_S$, $\Lambda \in \mbox{\fraktur g}$, is a vertical vector and $(\mbox{\bigscript R}_Y)_{\xi}$, $Y \in \mbox{\fraktur m}$, is a horizontal vector. These two vectors are orthogonal since, by Eq.(\ref{vorterm}),
\beqa
{\textbf g}_{\xi}((\mbox{\bigscript R}_\Lambda)_S, (\mbox{\bigscript R}_Y)_{\xi}) & = & {\textbf g}_{\xi}((\mbox{\bigscript R}_{R^t A \Lambda A^{-1} R})_\xi, (\mbox{\bigscript R}_Y)_{\xi}) \nonumber \\
& = & \mbox{tr}\left( R^t A \Lambda A^{-1} R (\xi \xi^t) Y^t \right) \nonumber \\
& = & \mbox{tr} \left( R^t A \Lambda A R Y \right) \nonumber \\
& = & -\mbox{tr} \left( Y^t R^t A \Lambda A R \right) \nonumber \\
& = & 0\,. 
\eeqa
In addition, observe that the sums of the dimensions of the vertical space and the horizontal space add to the dimension of the tangent space $T_\xi {\cal P}$. Therefore the tangent space is a direct sum of the horizontal and vertical subspaces. Moreover, if $(\mbox{\bigscript R}_Y)_{\xi}$ is a horizontal vector and $g\in G$, then right invariance implies
\beq
(R_g)_\ast (\mbox{\bigscript R}_Y)_{\xi} = (\mbox{\bigscript R}_Y)_{\xi g^{-1}} ,
\eeq
or $(R_g)_\ast H_{\xi} = H_{\xi g^{-1}}$. Since the assignment of the horizontal subspace $H_\xi$ is also smooth, it defines a connection $\Gamma$ on ${\cal P}$. The Christoffel symbols are found by determining explicitly the horizontal lifts of the vibrational and rotational vectors.

The vibrational vectors are horizontal since $Y = R^t A^{-1}\dot{A} R$ is a symmetric matrix, see Eq.(\ref{vibterm}). Therefore the Christoffel symbols for the vibrational vectors are zero. But the rotational vectors are not horizontal because
\beq
{\textbf g}_\xi ((\mbox{\bigscript R}_{\mbox{\fraktur e}_i})_{\! R}, (\mbox{\bigscript R}_{\mbox{\fraktur e}_b})_S) = \mbox{tr}\left( A \mbox{\fraktur e}_i A \mbox{\fraktur e}_b \right) = -2 \delta_{i b} a_j a_k \neq 0 
\eeq
for $i, j, k$ cyclic. Note that the inner product of two vertical vectors is also nonzero,
\beq
{\textbf g}_\xi ((\mbox{\bigscript R}_{\mbox{\fraktur e}_a})_S, (\mbox{\bigscript R}_{\mbox{\fraktur e}_b})_S) = -\mbox{tr}\left( A^2 \mbox{\fraktur e}_a \mbox{\fraktur e}_b \right) = \delta_{a b} (a_j^2 + a_k^2) 
\eeq
for $a, j, k$ cyclic. In order for $(\mbox{\bf E}_i)_\xi$ to be the horizontal lift of $(\mbox{\bigscript R}_{\mbox{\fraktur e}_i})_{\! R}$, it is necessary and sufficient that, for $b = 1, 2, 3\,$,
\beqa
0 & = & {\textbf g}_\xi ( (\mbox{\bf E}_i)_\xi , (\mbox{\bigscript R}_{\mbox{\fraktur e}_b})_S ) \nonumber \\
& = & {\textbf g}_\xi ( (\mbox{\bigscript R}_{\mbox{\fraktur e}_i})_{\! R} + \Gamma^a_i(q) (\mbox{\bigscript R}_{\mbox{\fraktur e}_a})_S  , (\mbox{\bigscript R}_{\mbox{\fraktur e}_b})_S )  \nonumber \\
& = & -2 \delta_{i b} a_j a_k + \Gamma^b_i(q) (a_j^2 + a_k^2).
\eeqa
The off-diagonal Christoffels for the rotational vectors vanish, and the diagonal values are
\beq
\Gamma_i^i (q) = \frac{2 a_j a_k}{(a_j^2 + a_k^2)} \qquad (i, j, k \;\;\mbox{cyclic}).
\eeq
Thus, the Riemannian connection for which the horizontal space is perpendicular to the vertical space corresponds to irrotational flow.

\subsection{Invariant Connection}
The falling cat connection is the {\em invariant} connection on the Lie group ${\cal P}$. Since {\fraktur g} is the algebra of antisymmetric matrices and {\fraktur m} is the vector space of symmetric matrices, the Lie algebra of the group ${\cal P}$ is a direct sum of vector spaces, $M_3(\mbox{\fields R}) = \mbox{\fraktur g} \oplus \mbox{\fraktur m}$. Moreover the vector space {\fraktur m} is invariant with respect to the adjoint group transformation, $Ad_g(\mbox{\fraktur m}) \subset \mbox{\fraktur m}$ for all $g\in G$. These two properties of {\fraktur m} are necessary and sufficient for
\beq
H_{\xi} = \left\{ (\mbox{\bigscript L}_Y)_{\xi} = -(\mbox{\bigscript R}_{Ad_\xi Y})_\xi \in T_{\xi}{\cal P} \ | \  Y \in \mbox{\fraktur m} \right\}
\eeq
to be a horizontal subspace \cite{KN}. In order to see that, note that the vertical vectors can be expressed in left invariant form,
\beq
V_{\xi} = \left\{ (\mbox{\bigscript R}_\Lambda)_{S} = -(\mbox{\bigscript L}_{S^t \Lambda S})_\xi \in T_{\xi}{\cal P} \ | \ \Lambda \in \mbox{\fraktur g} \right\} .
\eeq 
The tangent space to the bundle at $\xi$ is a direct sum of the horizontal and vertical subspaces, because every matrix is a linear combination of a symmetric matrix $Y$ and an antisymmetric matrix $S^t \Lambda S$. The right invariance of the horizontal subspaces is a consequence of
\beq
(R_g)_\ast (\mbox{\bigscript R}_{Ad_\xi Y})_\xi = (\mbox{\bigscript R}_{Ad_\xi Y})_{\xi g^{-1}} = (\mbox{\bigscript R}_{Ad_{\xi g^{-1}} Ad_g Y})_{\xi g^{-1}} \in H_{\xi g^{-1}},
\eeq
since $Ad_g Y \in \mbox{\fraktur m}$ for all $g\in G$ and $Y\in \mbox{\fraktur m}$. The assignment of the subspaces is smooth, so $H_{\xi}$ is indeed a horizontal subspace.

The relation
\beq
\dot{a}_i \left(\frac{\partial}{\partial a_i}\right)_A  = -(\mbox{\bigscript R}_{R^t A^{-1}\dot{A} R})_\xi  =  (\mbox{\bigscript L}_{S^t A^{-1}\dot{A} S})_\xi  
\eeq
shows that the vibrational vectors are horizontal ($S^t A^{-1}\dot{A} S$ is symmetric), but the rotational vectors are not horizontal since
\beq
(\mbox{\bigscript R}_{\mbox{\fraktur e}_i})_{\! R} = -(\mbox{\bigscript R}_{R^t \mbox{\fraktur e}_i R})_\xi = (\mbox{\bigscript L}_{S^t A^{-1} \mbox{\fraktur e}_i A S})_\xi
\eeq
and $S^t A^{-1} \mbox{\fraktur e}_i A S$ is not symmetric. If the matrix $S^t A^{-1} \mbox{\fraktur e}_i A S$ is expressed as a sum of symmetric $X_s$ and antisymmetric $X_a$ parts, i.e., $X_s = (S^t A^{-1} \mbox{\fraktur e}_i A S - S^t A \mbox{\fraktur e}_i A^{-1} S )/2$, $X_a =  (S^t A^{-1} \mbox{\fraktur e}_i A S + S^t A \mbox{\fraktur e}_i A^{-1} S )/2$, the angular momentum may be written as a sum of horizontal and vertical vectors 
\beq
(\mbox{\bigscript R}_{\mbox{\fraktur e}_i})_{\! R} = (\mbox{\bigscript L}_{X_s})_\xi + (\mbox{\bigscript L}_{X_a})_\xi \in H_\xi \oplus V_\xi.
\eeq 
The horizontal lifts of the angular momentum vectors are
\beqa
(\mbox{\bf E}_i)_\xi & = & (\mbox{\bigscript R}_{\mbox{\fraktur e}_i})_{\! R} + \Gamma^a_i(q) (\mbox{\bigscript R}_{\mbox{\fraktur e}_a})_S \nonumber \\
& = &  (\mbox{\bigscript L}_{X_s})_\xi + \left( (\mbox{\bigscript L}_{X_a})_\xi -  \Gamma^a_i(q) (\mbox{\bigscript L}_{S^t \mbox{\fraktur e}_a S})_\xi \right) ,
\eeqa
where $(\mbox{\bigscript L}_{X_s})_\xi$ is the horizontal lift and the two vertical vectors in the parentheses must cancel. Therefore, the invariant connection is given by
\beq
\Gamma^a_i(q)\mbox{\fraktur e}_a = (A^{-1} \mbox{\fraktur e}_i A  +  A \mbox{\fraktur e}_i A^{-1} )/2,
\eeq
or the Christoffel symbols are diagonal and
\beq
\Gamma_i^i (q) = \frac{(a_j^2 + a_k^2)}{2 a_j a_k} ,
\eeq
where $i,j,k$ are cyclic.

\subsection{Curvature}
Given a bundle connection $\Gamma$, define a $1$-form $\omega$ on ${\cal P}$ with values in the Lie algebra {\fraktur g} of the structure group $G$ as follows: if $X\in T_\xi {\cal P}$ is a tangent vector in the bundle and $Z$ is its vertical component, then there is a unique element $\Lambda \in$ {\fraktur g} such that the vertical component is the fundamental vector field $\Lambda^\ast = Z$. At $\xi \in {\cal P}$, define $\omega(X) = \Lambda$. The form $\omega$ is called the {\em connection form} of the connection $\Gamma$. The connection form satisfies three general properties:
\[
\begin{array}{lll}
(1) & \omega(X) = 0, & \mbox{\ if and only if\ } X \mbox{\ is horizontal;} \\
(2) & \omega(\Lambda^\ast) = \Lambda & \mbox{\ for every\ } \Lambda\in \mbox{\fraktur g} ; \\
(3) & \omega((R_g)_\ast X) = g\, \omega(X)\, g^{-1} & \mbox{ for every\ } g\in G \mbox{\ and every vector field\ } X.
\end{array}
\]
The first two conditions are an immediate consequence of the definition of the connection $1$-form. The third property is proven by considering separately horizontal and vertical vectors. If $X$ is horizontal, then $(R_g)_\ast X$ is also horizontal and both sides in property three are zero. If $X$ is a vertical vector at $\xi \in {\cal P}$, then it equals some fundamental vector field, $X_\xi = (\Lambda^\ast)_\xi$ and the third property is a consequence of Eq.(\ref{adjoint1}). Conversely, if the second and third conditions are satisfied by some $1$-form $\omega$, then it defines a connection for which the horizontal subspaces are the vectors $X$ satisfying $\omega(X) = 0$.

If $\omega$ is the connection $1$-form, then its exterior derivative $d\omega$ is a $2$-form, i.e., a bilinear, antisymmetric form defined on pairs of tangent vectors $X, Y$ to the bundle by
\beq
d\omega (X, Y) = \hf \left\{ X(\omega(Y)) - Y(\omega(X)) - \omega([X, Y])  \right\} .
\eeq 
The {\em curvature form} $\Omega$ of $\omega$ is defined by $\Omega = (d\omega) h$, 
\beq
\Omega (X, Y) = d\omega (hX, hY) ,
\eeq
where $hX$ and $hY$ are the horizontal components of the vectors $X$ and $Y$, respectively.
If either $X$ or $Y$ is vertical, then $\Omega(X, Y) = 0$. If both $X$ and $Y$ are horizontal, then $2\Omega(X, Y) = - \omega([X, Y])$.

The curvature form of the Riemann ellipsoid bundle is, for $1\leq i,j \leq 3$, as follows:
\beqa
\Omega(\frac{\partial}{\partial a_i}, \frac{\partial}{\partial a_j} ) & = & 0  \mathletter{a} \\
\Omega(\frac{\partial}{\partial a_i}, \mbox{\bf E}_j ) & = &  \mbox{\fraktur T}_{ij}\, (Ad_{S^{-1}}\mbox{\fraktur e}_j) \mathletter{b} \\
\Omega(\mbox{\bf E}_i, \mbox{\bf E}_j ) & = &  \varepsilon_{ijk}  \mbox{\fraktur R}_{k}\, (Ad_{S^{-1}}\mbox{\fraktur e}_k),  \mathletter{c}
\eeqa
where the field tensors are given in terms of the connection by
\beqa
\mbox{\fraktur T}_{ij} & = & \half \frac{\partial \Gamma^j_j}{\partial a_i}  \mathletter{a} \\
\mbox{\fraktur R}_{k} & = & \half(\Gamma_k^k - \Gamma_i^i \Gamma_j^j) \qquad (i, j, k \;\;\mbox{cyclic}). \mathletter{b} \label{fieldtensors}
\eeqa
To derive the curvature expression, use the commutation relations among the horizontal vectors:
\beqa
\left[ \frac{\partial}{\partial a_i} , \frac{\partial}{\partial a_j} \right] & = & 0  \mathletter{a} \\
\left[ \frac{\partial}{\partial a_i} , \mbox{\bf E}_{j} \right]  & = &  - \frac{\partial \Gamma^j_j}{\partial a_i}  (Ad_{S^{-1}}\mbox{\fraktur e}_j)^\ast  \mathletter{b} \\
\left[ \mbox{\bf E}_i , \mbox{\bf E}_j \right]  & = & -\varepsilon_{ijk} \left( \mbox{\bf E}_k + (\Gamma_k^k - \Gamma_i^i \Gamma_j^j) (Ad_{S^{-1}}\mbox{\fraktur e}_k)^\ast \right) .  \mathletter{c}
\eeqa

In general, if $\theta$ is any {\fraktur g}-valued $p$-form, then its {\em covariant derivative} $D\theta$ is a {\fraktur g}-valued $(p+1)$-form given by the composition of horizontal projection and exterior differentiation, $D\theta = (d\theta)h$. The covariant derivative $D\theta$ and the exterior derivative $d\theta$ are totally antisymmetric in their arguments that are $p+1$ tangent vectors to the bundle. The curvature form is the covariant derivative of the connection $1$-form, $\Omega = D\omega$. The Bianchi identity states that the covariant derivative of the curvature form is zero, $D\Omega = 0$. For any three horizontal vectors $X, Y, Z$, the covariant derivative of the curvature form is given by
\beqa
3\, D\Omega (X, Y, Z) & = & X (\Omega(Y, Z)) + Y (\Omega(Z, X)) + Z (\Omega(X, Y)) \nonumber \\ 
& & - \Omega([X, Y], Z) - \Omega([Y, Z], X) - \Omega([Z, X], Y)\, .\label{threeform}
\eeqa
If all three horizontal vectors are vibrational, then it is clear that the covariant derivative is zero. Although not self-evident, the covariant derivative is also trivially zero if the three horizontal vectors are rotational. If two arguments are vibrational and one rotational, then
\beq
0 = 3\, D\Omega (\mbox{\bf E}_i, \frac{\partial}{\partial a_j}, \frac{\partial}{\partial a_k})  =  \half \left( \frac{\partial \mbox{\fraktur T}_{ki}}{\partial a_j} - \frac{\partial \mbox{\fraktur T}_{ji}}{\partial a_k}  \right) (Ad_{S^{-1}}\mbox{\fraktur e}_i)  \label{bianchi1}
\eeq
for all $1\leq i, j, k \leq 3$. If one argument is vibrational and the other two rotational, then 
\beq
0 = 3\, D\Omega (\mbox{\bf E}_i, \mbox{\bf E}_j, \frac{\partial}{\partial a_k})  = \varepsilon_{ijm} \left( \Gamma^i_i \mbox{\fraktur T}_{kj} + \Gamma^j_j \mbox{\fraktur T}_{ki} - \mbox{\fraktur T}_{km} + \frac{\partial\mbox{\bigscript R}_{m}}{\partial a_k}  \right) (Ad_{S^{-1}}\mbox{\fraktur e}_m) \label{bianchi2}. 
\eeq
The latter equation is proven using the relations
\beqa
[ \mbox{\fraktur e}_i ,\mbox{\fraktur e}_j ] & = & -\varepsilon_{ijk} \mbox{\fraktur e}_k  \mathletter{a} \\ 
(\mbox{\bigscript R}_{\mbox{\fraktur e}_i})_S  (Ad_{S^{-1}}\mbox{\fraktur e}_j) & = &  Ad_{S^{-1}}[ \mbox{\fraktur e}_i ,\mbox{\fraktur e}_j ] . \mathletter{b}
\eeqa

Conversely, suppose the Bianchi identities, Eqs.(\ref{bianchi1},\ref{bianchi2}), are satisfied by the field tensors. Define three ``displacement vectors'', $\mathbf{D}_i = (\mbox{\fraktur T}_{1i} ,\mbox{\fraktur T}_{2i} ,\mbox{\fraktur T}_{3i} )$, i.e., the $i^{th}$ column of the tensor is the vector $\mathbf{D}_i$. The first Bianchi identity, Eq.(\ref{bianchi1}), is equivalent to the vanishing of the curl of each displacement vector 
\beq
\mathbf{\bbf{\nabla}}_a \times \mathbf{D}_i = 0 .
\eeq
Since ${\cal A}$ is a simply connected region, each irrotational displacement vector is the gradient of some ``electric potential'', 
\beq
\mathbf{D}_i = \half \, \mathbf{\bbf{\nabla}}_a \phi_i .
\eeq
By substituting the potentials $\phi_i$ for {\fraktur T}, the second Bianchi identity, Eq.(\ref{bianchi2}), simplifies to
\beqa
0 & =&  \varepsilon_{ijm} \left\{ \frac{\partial}{\partial a_k} ( 2\mbox{\fraktur R}_{m} - \phi_m +\phi_i \phi_j) + (\Gamma^i_i - \phi_i)  \frac{\partial \phi_j}{\partial a_k} +  (\Gamma^j_j - \phi_j)  \frac{\partial \phi_i}{\partial a_k}  \right\} \nonumber \\
& &  \hskip0.2in \times (Ad_{S^{-1}}\mbox{\fraktur e}_m) .
\eeqa   
If the potentials are set equal to the Christoffel symbols, $\phi_i = \Gamma^i_i$, then the above equation determines, up to additive constants, the {\fraktur R} tensor.

\subsection{Projection to the Base Manifold}
The connection one-form $\omega$ and the curvature two-form $\Omega$ on the bundle ${\cal P}$ can be projected down to the base manifold. The projection depends on the choice of a local trivialization $\tau : {\cal Q}\longrightarrow {\cal P}$. If $\theta$ is any {\fraktur g}-valued $p$-form on the bundle, then its {\em pullback} $\tau^\ast \theta$ is a {\fraktur g}-valued $p$-form on the base manifold. The pullback of the connection one-form $\mbox{\bigscript A} = \tau^\ast \omega$ is a one-form on ${\cal Q}$ given by
\beq
\mbox{\bigscript A}_q (X) = \omega(\tau_\ast X) \mbox{\ for any tangent vector\ } X \in T_q({\cal Q}) .
\eeq
For our usual choices of a local trivialization, $\tau(q) = R^t A$, and an adapted basis of tangent vectors, $\{({\bf e}_i)_q, 1\leq i\leq 6\}$, the pullback is just
\beqa
\mbox{\bigscript A} (\frac{\partial}{\partial a_k}) & = & 0 \mathletter{a} \\
\mbox{\bigscript A} (\mbox{\bigscript R}_{\mbox{\fraktur e}_k}) & = & \Gamma^k_k(q) \mbox{\fraktur e}_k .\mathletter{b} 
\eeqa
If the pullback is applied to the tangent vector of a curve in the base manifold, then it produces the vortex velocity
\beq
\mbox{\bigscript A} (V(t)) = - \omega_k \Gamma^k_k \mbox{\fraktur e}_k = -\Lambda(t) . 
\eeq

The pullback of the curvature is defined similarly, $F = \tau^* \Omega$, and it is a two-form on the base manifold:
\beq
F_q (X, Y) = \Omega(\tau_\ast X, \tau_\ast Y) \mbox{\ for all tangent vectors\ } X, Y \in T_q({\cal Q}) .
\eeq
The tensor $F$ is evaluated to be
\beqa
F( \frac{\partial}{\partial a_i}, \frac{\partial}{\partial a_j} ) & = & 0 \mathletter{a} \\
F(\frac{\partial}{\partial a_i} , \mbox{\bigscript R}_{\mbox{\fraktur e}_j}  ) & = & \mbox{\fraktur T}_{ij} \mbox{\fraktur e}_j  \mathletter{b} \\
F(\mbox{\bigscript R}_{\mbox{\fraktur e}_i} , \mbox{\bigscript R}_{\mbox{\fraktur e}_j} ) & = & \varepsilon_{ijk}\mbox{\fraktur R}_k \mbox{\fraktur e}_k . \mathletter{c} 
\eeqa

The Bianchi identity on the bundle takes a slightly more complicated expression on the base manifold
\beq
dF + [ \mbox{\bigscript A}, F ] = 0 , \label{Cartanform}
\eeq 
where $dF$ denotes the exterior derivative of $F$, see Eq.(\ref{threeform}), and the commutator is the three-form defined by
\beqa
3\, [ \mbox{\bigscript A}, F ] (X, Y, Z) & = & [ \mbox{\bigscript A}(X) , F(Y, Z) ] +  [ \mbox{\bigscript A}(Y) , F(Z, X) ] +  [ \mbox{\bigscript A}(Z) , F(X, Y) ] \nonumber \\
& &  
\eeqa
for all tangent vectors $X, Y, Z \in T_q({\cal Q})$.The identity Eq.(\ref{Cartanform}) is verified by letting $X, Y$, and $Z$ run over the basis of the tangent space.

\section{CONCLUSIONS}

The connections corresponding to rigid rotation, irrotational flow, and the falling cat were shown to be natural geometrical or group-theoretical concepts. Although not mathematically natural, other choices of Christoffel symbols define nonholonomic constraint forces that are not excluded by physical law. For example, the $S$-type Riemann ellipsoids are a sequence of special case solutions for which the angular momentum, Kelvin circulation, and the angular and vortex velocity vectors are aligned with a principal axis, say the $1$-axis \cite{LEB, Chandrasekhar69}. This sequence is indexed by a continuous real parameter $f$ restricted to the interval $[-2, 0]$. There is only one horizontal lift to consider and the Christoffel symbol is given by
\beq
\Gamma_1^1 (q) = -\frac{f a_2 a_3}{(a_2^2 + a_3^2)} .
\eeq
At $f = 0$, the connection yields rigid rotation, and, at $f = -2$, it is irrotational flow. The $S$-type ellipsoids are the simplest models that allow for a continuous interpolation between rigid rotation and irrotational flow. This connection has no natural geometrical or group-theoretic significance -- but it does model a variety of rotating physical systems.

The unity found in rotating systems is remarkable \cite{WJS74, Dankova}. Besides astrophysical systems \cite{LEB, Chandrasekhar69}, the Riemann ellipsoid model applies to classical liquid droplets \cite{Rosenkilde} and quantum atomic nuclei \cite{Rose88, Garrett, Rose92c, GOO94}. For example, the kinetic energy of a Riemann ellipsoid, Eq.(\ref{Riemannkinetic}), can be derived from self-consistency of a mean field approximation to the quantum fermion system of rotating nucleons \cite{Rose92b}. 

A basic science problem is to determine the connection from the interactions among the particles that form a rotating system. The horizontal lift of a curve describing the rotation of a body depends on the coupling between the collective degrees of freedom of the Riemann ellipsoidal model and those of the individual particles. Equations that incorporate these interactions into the gauge theory and whose unique solution are the Christoffel symbols are required for a complete theory of collective rotation. They must involve a coordinate independent object and the curvature form is the obvious candidate. The Bianchi identity, $D\Omega = 0$, partially determines the Christoffel symbols, but it is not sufficient. There must be another equation that relates the bundle curvature to the microscopic physics.

The analogy with classical theory of electricity and magnetism is helpful. This is an Abelian gauge theory  with space-time as the base manifold and $U(1)$ as the structure group. The curvature form is the Faraday tensor and the Bianchi identity is equivalent to two of Maxwell's equations, $\bbf{\nabla} \cdot \mathbf{B} =0$ and $\bbf{\nabla}\times \mathbf{E} + \partial\mathbf{B}/\partial t =0$. The other two of Maxwell's equations involve the current and are equivalent to a second independent equation for the Faraday tensor. It is the analogous second equation for the gauge theory of rotating Riemann ellipsoids that is unknown.

The quantization of a gauge theory is achieved by constructing associated bundles. For each irreducible representation of the structure group $G \simeq SO(3)$, an inequivalent associated bundle and a different quantization can be defined. In a subsequent article, the associated bundle theory will be presented. The quantized gauge theory applies directly to nuclear structure physics, e.g., the liquid drop model \cite{Bohr52} is the associated bundle constructed from the one-dimensional representation of the structure group. Although the quantization of the Riemann model was considered previously \cite{ros79, ros93c}, the differential structure of the quantized theory has not been studied yet.

\begin{acknowledgment}
The authors would like to thank the Institute of Nuclear Physics and the organizers, F. Iachello and J. Ginocchio, of the program entitled, ``Algebraic Methods in Many-Body Physics," for their support. Valuable discussions with J.\ Bryan, Ts.\ Dankova, and E.\ Ihrig contributed to this work.
\end{acknowledgment}

\newpage
\begin{figure}[htbp]\centering
\mbox{\psfig{figure=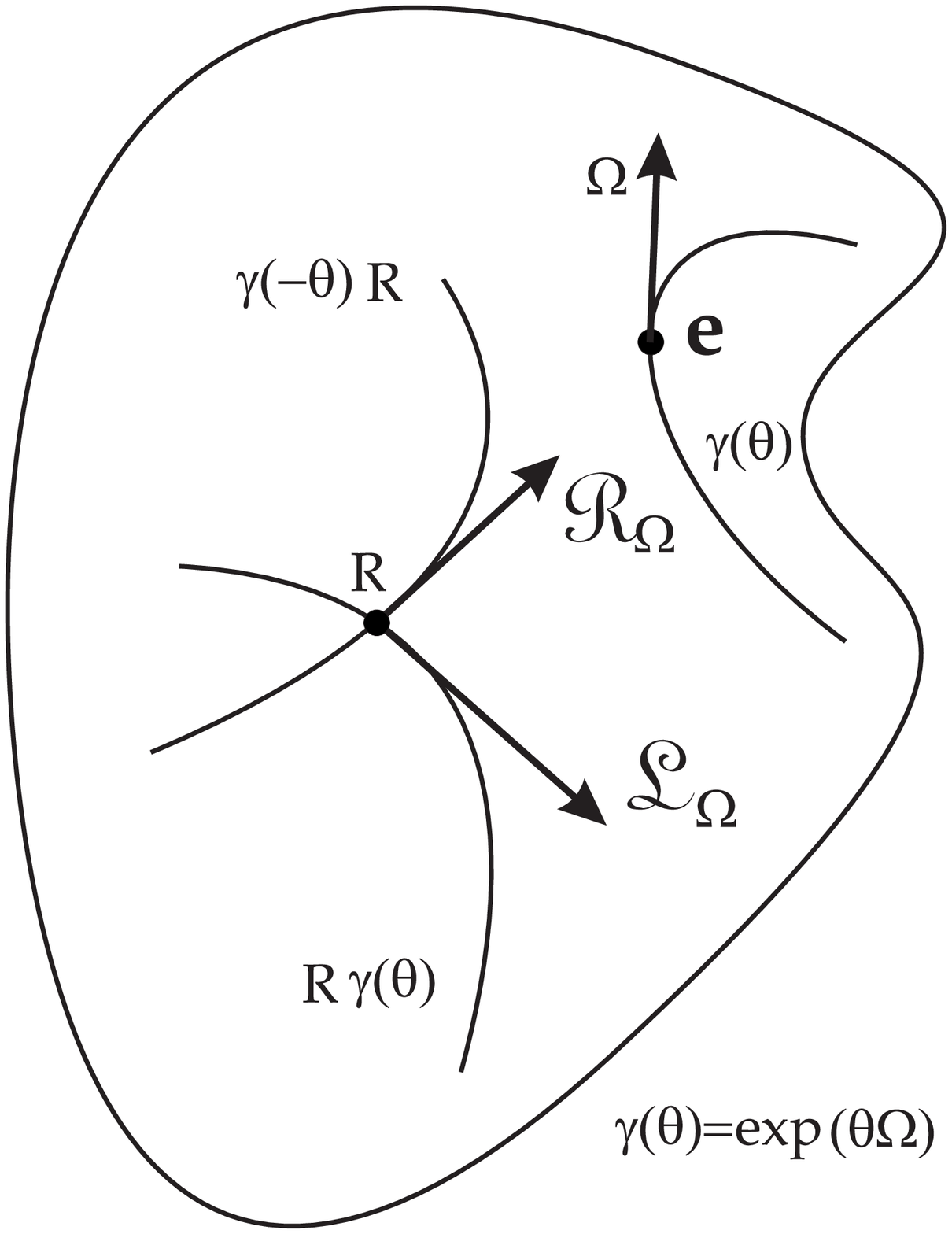,height=3in,width=3in,clip=}}
\caption{Left invariant {\scriptcaption L}$_\Omega$ and right invariant {\scriptcaption R}$_\Omega$ vector fields.}
\end{figure}

\begin{figure}[htbp]\centering
\mbox{\psfig{figure=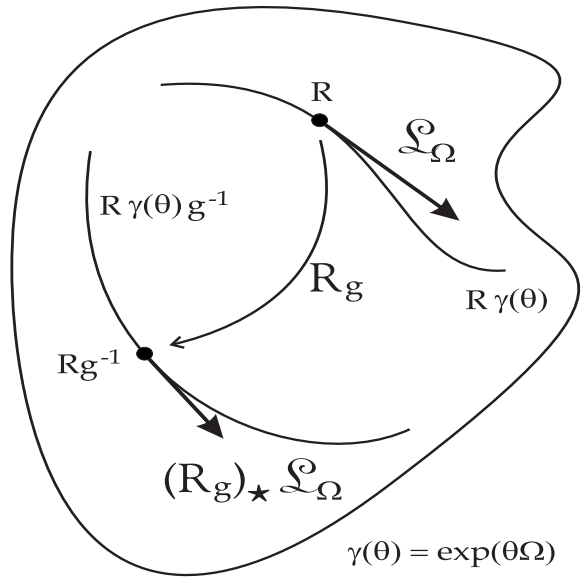,height=3in,width=3in,clip=}}
\caption{Right translation $(R_g)_\ast$ of a left invariant vector field {\bigscript L}$_\Omega$.}
\end{figure}

\begin{figure}[htbp]\centering
\mbox{\psfig{figure=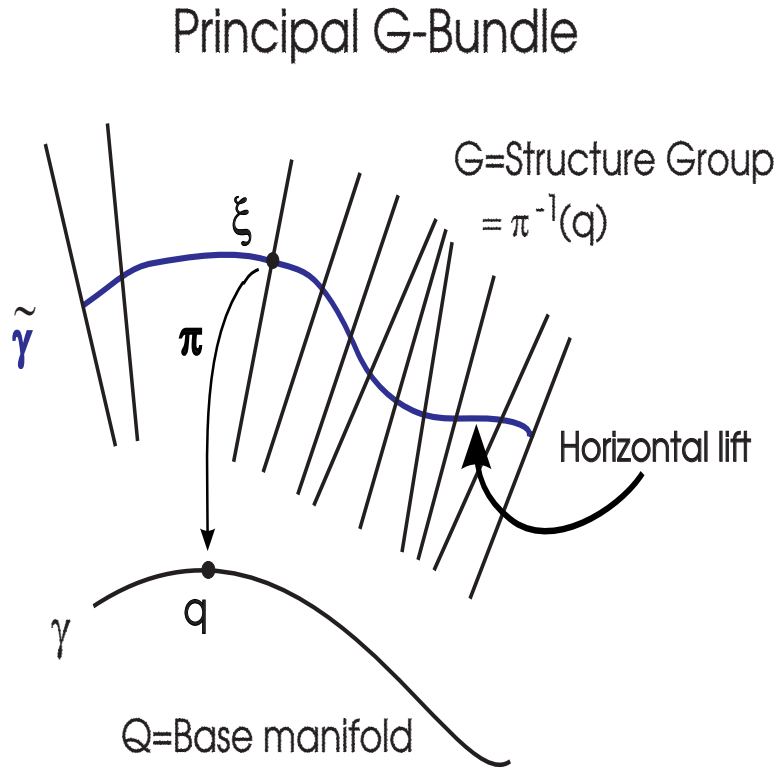,height=4in,width=4in,clip=}}
\caption{A curve $\gamma$ in the base manifold lifts to a curve $\tilde{
\gamma}$ in the bundle. The curve in the bundle is called the horizontal lift. It depends uniquely on the bundle connection.}
\end{figure}


\end{article}
\end{document}